\RequirePackage{fix-cm}
\documentclass{article}
\usepackage{iclr2027_conference,times}
\iclrfinalcopy
\usepackage[T1]{fontenc}
\usepackage[utf8]{inputenc}
\usepackage{algorithm2e}
\usepackage{algpseudocode}
\usepackage{amsmath}
\usepackage{amsthm}
\usepackage{amssymb}
\usepackage{thmtools}
\usepackage{bbm}
\usepackage{esint}
\usepackage{xcolor}
\usepackage{bm}
\usepackage{float} 
\usepackage{comment}
\usepackage{microtype}
\usepackage[unicode=true,
 bookmarks=false,
 breaklinks=true,pdfborder={0 0 0},pdfborderstyle={},backref=section,colorlinks]
 {hyperref}
 \hypersetup{citecolor=[RGB]{119,185,0}}

\makeatletter
\usepackage{changebar}
\providecommand{\lyxadded}[3]{}

\renewcommand{\lyxadded}[3]{
  {\protect\cbstart\color{lyxadded}{}#3\protect\cbend}
}

\usepackage{graphicx}
\usepackage{amsfonts}
\usepackage{amsthm}
\usepackage{mathtools}
\usepackage{verbatim}
\usepackage{multicol}
\usepackage{array}
\usepackage{tabularx}
\usepackage{ragged2e}
\newcolumntype{P}[1]{>{\RaggedRight\arraybackslash}p{#1}}

\usepackage{booktabs}

\usepackage{cleveref}
\usepackage{titlesec}
\titlespacing*{\paragraph}{0pt}{0em}{0.5em}

\usepackage{footmisc}
\usepackage{enumitem}
\usepackage{xpatch}
\usepackage{tikz}
\usepackage{hyperref}
\makeatletter
  {\begin{titlepage}\def\@thanks{}}%
  {\end{titlepage}}
\xpatchcmd\titlepage{\setcounter{page}\@ne}{}{}{}
\xpatchcmd\endtitlepage{\setcounter{page}\@ne}{}{}{}

\newcommand{\probP}{\text{I\kern-0.15em P}}

\theoremstyle{remark}
\newtheorem{lemma}{Lemma}
\newtheorem{corollary}{Corollary}

\newtheorem{thm}{Theorem}
\newtheorem{assump}{Assumption}
\newtheorem{remark}{Remark}

\LinesNumbered
\RestyleAlgo{algoruled} 

\makeatother

\definecolor{adbskyyellow}{HTML}{F2D22E}
\definecolor{PCAcolor}{cmyk}{0, 0.85, 0.85, 0.50}
\renewcommand\emph[1]{{\color{PCAcolor}\textit{#1}}}

\usepackage[most]{tcolorbox}

\tcbset{
  enhanced,
  breakable, 
}
\definecolor{tufteframe}{RGB}{120,120,120}
\definecolor{tufteback}{gray}{0.99}

\newtcolorbox{colorboxblue}{
  enhanced, breakable,
  colback=white,
  frame hidden, borderline west={1.2pt}{0pt}{black!40},
  left=2mm, right=2mm, top=1mm, bottom=1mm,
  boxrule=0pt, 
  width=\linewidth,     
  before skip=10pt, after skip=10pt,
  colframe=NavyBlue, colback=SkyBlue!8
}

\newtcolorbox{colorboxbrown}{
  enhanced, breakable,
  colback=white,
  frame hidden, borderline west={1.2pt}{0pt}{black!40},
  left=2mm, right=2mm, top=1mm, bottom=1mm,
  boxrule=0pt, 
  width=\linewidth,     
  before skip=10pt, after skip=10pt,
  colframe=RawSienna, colback=RawSienna!5
}

\newtcolorbox{colorboxyellow}{
  enhanced, breakable,
  colback=white,
  frame hidden, borderline west={2.2pt}{1pt}{black!40},
  left=2mm, right=2mm, top=1mm, bottom=1mm,
  boxrule=0pt, 
  width=\linewidth,     
  before skip=5pt, after skip=5pt,
  colframe=adbskyyellow, colback=white!20
}

\begin{document}
\title{Power-Optimal Risk Control via Standardized Margin Calibration}

\author{
\parbox{0.95\textwidth}{
\centering
Ruinan Wang\textsuperscript{1},
Tian Bai\textsuperscript{2},
Yueming Shen\textsuperscript{3},\\[0.3em]
Xiang Yu\textsuperscript{4},
Ying Yang\textsuperscript{5},
Archer Y. Yang\textsuperscript{3,6}
}
}

\maketitle

\fancyhead{}
\renewcommand{\headrulewidth}{0pt}

\begingroup
\renewcommand{\thefootnote}{}
\footnotetext{%
\textsuperscript{1}Department of Mathematical Sciences,
Tsinghua University, Beijing, China.
\textsuperscript{2}Department of Statistics,
Stanford University, CA, USA.
\textsuperscript{3}Department of Mathematics and Statistics,
McGill University, Montreal, Canada.
\textsuperscript{4}Eli Lilly and Company,
Indianapolis, IN, USA.
\textsuperscript{5}Department of Statistics and Data Science,
Tsinghua University, Beijing, China.
\textsuperscript{6}Mila -- Quebec AI Institute,
Montreal, Quebec, Canada.
Correspondence to: Archer Y. Yang
\textless\href{mailto:archer.yang.yi@gmail.com}
{\nolinkurl{archer.yang.yi@gmail.com}}\textgreater.
}
\endgroup

\fancyhead{}
\renewcommand{\headrulewidth}{0pt}

\begin{abstract}
Machine learning can screen far more candidates than can be experimentally validated, creating a need for selection procedures that identify promising candidates while controlling risk before costly experimental validation. We propose Uncertainty-Normalized Empirical Calibration (UNEC), a $p$-value-free framework that ranks candidates by uncertainty-normalized predictive margins and directly calibrates a selection cutoff using held-out losses. UNEC supports flexible prediction models, general bounded losses, and diverse selection targets. We study marginal discovery risk (MDR) and selective discovery risk (SDR), with SDR reducing to the false discovery rate under binary loss. We establish finite-sample MDR control for an adjusted rule and asymptotic SDR control with explicit limiting risk and power. Under a location--scale model with a common one-sided binary target, we further show that UNEC asymptotically attains the oracle power benchmark among observation-wise selection rules. Simulations and molecular screening applications in solubility, toxicity, and lipophilicity demonstrate favorable risk--power trade-offs and substantial computational gains across binary and graded-loss settings.
\end{abstract}

\section{Introduction}
\label{sec:introduction}

Machine learning increasingly guides decisions about which candidates deserve further investigation and which predictions can be acted upon. Applications include prioritizing antibacterial compounds for experimental testing \citep{stokes2020deep}, screening generated proteins for desired biological activity \citep{madani2023large}, identifying stable crystal structures \citep{merchant2023scaling}, selecting patients likely to benefit from treatment \citep{lei2021conformal}, and accepting trustworthy outputs from large language models \citep{gui2024conformal}. These decisions must be made before the relevant outcomes or quality assessments are available, while unsuitable selections can waste resources or compromise downstream decisions.

This motivates the problem of \emph{risk-controlled selection}: selecting promising observations while controlling a prespecified measure of selection risk. Much of the existing literature focuses on binary false-selection errors. Conformal selection (CS), for example, combines conformal $p$-values with multiple-testing procedures to control the false discovery rate (FDR) \citep{jin2023selection}. In many applications, however, the cost or extent of an error matters. Drug candidates differ in synthesis cost \citep{fromer2024algorithmic}, so counting inactive selections alone does not capture the resources wasted. Likewise, generated responses may contain different proportions of unsupported factual claims \citep{min2023factscore}. These considerations motivate general bounded losses that quantify costs or graded errors beyond binary false-selection indicators.

Recent work such as SCoRE provides finite-sample selective risk control for general bounded losses through conformal $e$-values \citep{bai2026csgeneralrisk}. We pursue a complementary direction: can risk-controlled selection be achieved directly from predictions and held-out losses, without constructing conformal $p$- or $e$-values? Such an alternative is particularly attractive if it remains flexible and computationally efficient across different losses and selection targets.

We propose \emph{uncertainty-normalized empirical calibration} (UNEC), a simple $p$-value-free framework combining standardized-margin ranking with direct empirical risk calibration. By normalizing predicted target margins by local uncertainty, UNEC provides a flexible construction supporting general bounded losses, observation-specific thresholds, and interval targets. Calibration uses sorted scores and cumulative losses, offering substantial computational savings over the evaluated SCoRE implementations. Our theory establishes finite-sample marginal discovery risk (MDR) control for an adjusted rule and asymptotic selective discovery risk (SDR) control under suitable conditions. For binary loss with a common one-sided target, we further establish asymptotic power optimality under a location--scale model and appropriate estimation conditions. Simulations and molecular screening applications demonstrate the effectiveness of UNEC, with competitive risk--power performance and substantial computational gains.

\section{Background and Related Work} \label{sec:setup}
\paragraph{Problem setup}
Suppose we observe a labeled dataset $\mathcal D_{\mathrm{obs}}=\{(X_i,Y_i)\}_{i=1}^n$ with index set $I_{\mathrm{obs}}$, where $X_i\in\mathcal X\subseteq\mathbb R^p$ denotes the covariates and $Y_i\in\mathcal Y\subseteq\mathbb R$ denotes the response. In addition, we are given an unlabeled test set $\mathcal D_{\mathrm{test}}=\{X_{n+j}\}_{j=1}^m$ with index set $I_{\mathrm{test}}$, where the responses $Y_{n+1},\ldots,Y_{n+m}$ are unobserved at the time of selection. Throughout the paper, we assume that the observed and test samples are independent and identically distributed (i.i.d.) copies of a random pair $(X,Y)$, unless otherwise specified.

For each test observation \(j\), the selection target is represented by a possibly random target descriptor \(T_j\), observed before selection, which determines a null region \(\mathcal R_{0,j}=\mathcal R_0(T_j)\subseteq\mathcal Y\) and a target region \(\mathcal R_{1,j}=\mathcal Y \setminus \mathcal{R}_{0,j}\). The corresponding random hypothesis is
$$
H_{0j}:Y_{n+j}\in\mathcal R_{0,j},
\qquad
H_{1j}:Y_{n+j}\in\mathcal R_{1,j},
\qquad j=1,\ldots,m.
$$
For example, in the one-sided setting with larger responses being preferable, one may take \(T_j=c_j\) and \(\mathcal R_{0,j}=(-\infty,c_j),\ \mathcal R_{1,j}=[c_j,\infty).\) In a two-sided setting, one may take \(T_j=(a_j,b_j)\), and define the target region as the interval \([a_j,b_j]\).

Since $Y_{n+j}$ is unavailable at the testing stage, our goal is to construct an estimated selection set $\widehat{S}\subseteq\{1,\ldots,m\}$ using only the observed data $\mathcal D_{\mathrm{obs}}$, the test covariates $\{X_{n+j}\}_{j=1}^m$, and the target descriptors $\{T_j\}_{j=1}^m$. To quantify the risk incurred by selecting a test observation, we consider a known loss function $\ell:\mathcal X\times\mathcal Y\times\mathcal T\to[0,1]$, where $\mathcal T$ denotes the space of possible target descriptors. For any \(i \in I_{\mathrm{obs}} \cup I_{\mathrm{test}}\) and any \(j \in I_{\mathrm{test}}\), define the bounded loss as  $L_{i,j}=\ell(X_{i},Y_{i},T_j)\in[0,1].$ Generally, $L_{i,j}$ can be either a discrete or continuous bounded loss, such as a false selection loss weighing the cost or a distance from the target region. 

Let $a_j=\mathbbm{1}\{j\in\widehat S\}$ denote the selection decision for the $j$-th test observation, where $a_j=1$ means that the observation is selected and $a_j=0$ means that it is not selected.
Following the marginal and selective deployment risks introduced by \citet{bai2026csgeneralrisk}, we consider two analogous risk criteria in our selection setting: marginal discovery risk (MDR) and selective discovery risk (SDR), defined as
$$
\mathrm{MDR}=\frac{1}{m}\sum_{j=1}^m\mathbb E\left[L_{n+j,j}a_j\right], \qquad
\mathrm{SDR}=\mathbb E\left[\frac{\sum_{j=1}^m L_{n+j,j}a_j}{1\vee \sum_{j=1}^m a_j}\right].
$$
MDR measures the average marginal risk incurred over the test population, while SDR measures the expected average loss among selected observations. Under the binary loss $L_{n+j,j}=\mathbbm{1}\{Y_{n+j}\in\mathcal R_{0,j}\}$, the SDR reduces to the false discovery rate (FDR).

In addition to risk control, we aim to select as many desirable observations as possible. We define the power as
$$\mathrm{Power}=\mathbb E\left[\frac{\sum_{j=1}^m \mathbbm{1}\{Y_{n+j}\in\mathcal R_{1,j}\} \cdot a_j}{1\vee \sum_{j=1}^m \mathbbm{1}\{Y_{n+j}\in\mathcal R_{1,j}\}}\right].$$
Given a target risk level $q\in(0,1)$, our objective is to construct a data-driven selection rule $\widehat{S}$ such that either $\mathrm{MDR}\le q$ or $\mathrm{SDR}\le q$, while achieving high power.

\paragraph{Conformal selection}
Conformal selection provides a distribution-free route to selection with FDR control by combining conformal scores with multiple testing \citep{jin2023selection}. Subsequent work has extended this framework to settings with covariate shift \citep{jin2026weighted}, multivariate responses \citep{bai2025multivariate}, and multiple simultaneous selection criteria \citep{hao2026multicondition}. These developments primarily expand the structure of the selection problem while retaining a binary notion of false selection. A different line of work studies power more directly. In particular, \citet{qin2025revamping} develops a Neyman--Pearson (NP) perspective in which candidates are ranked by evidence for exceeding the target and the selection threshold is calibrated to control FDR. UNEC is closer to this ranking-and-calibration viewpoint than to conformal multiple testing, but replaces likelihood-ratio-style scoring with an uncertainty-normalized margin learned from conditional location and scale.

\paragraph{Risk control}
A separate literature studies predictive procedures under more general notions of risk. Risk-controlling prediction sets \citep{bates2021distribution} and Learn then Test \citep{angelopoulos2025learn} obtain high-probability guarantees through concentration inequalities and hypothesis testing, respectively, while conformal risk control (CRC) targets expected bounded loss under monotonicity of the loss in a calibration parameter \citep{angelopoulos2024conformal}. These methods calibrate predictive objects or tuning parameters rather than a data-dependent selected subset. Most closely related to our setting, SCoRE extends selective prediction beyond binary errors by using risk-adjusted conformal $e$-values to control general bounded losses \citep{bai2026csgeneralrisk}. Relative to this line of work, UNEC studies empirical threshold calibration on an uncertainty-based ranking, leading to a different trade-off between finite-sample guarantees, asymptotic selective-risk control, and computational cost.

\section{Uncertainty-Normalized Empirical Calibration (UNEC)} \label{sec:method}

We propose a general framework termed Uncertainty-Normalized Empirical Calibration (UNEC) for risk-controlled selection, which combines uncertainty-normalized ranking with direct empirical MDR or SDR calibration. We develop the method primarily in the one-sided setting with a common threshold $c$, which captures the essential idea of the proposed procedure, and then briefly summarize extensions to observation-specific thresholds, conservative calibration, and two-sided targets, with further details deferred to Appendix~\ref{app:varying_thresholds}.

We focus on the one-sided selection problem with a common threshold $c$, corresponding to \(T_j=c,\ \mathcal R_{0,j}=(-\infty,c),\ \mathcal R_{1,j}=[c,\infty),\ j \in I_{\mathrm{test}}\) and $L_{i,1}=\cdots=L_{i,m}:=L_i, i \in I_{\mathrm{obs}}.$ UNEC uses sample splitting to construct and calibrate an uncertainty-normalized score. Specifically, we randomly split the observed dataset $\mathcal{D}_\mathrm{obs}$ into three disjoint subsets $\mathcal{D}_{\mathrm{train1}},$ $\mathcal{D}_{\mathrm{train2}}$ and $\mathcal{D}_{\mathrm{cal}}$ of sizes \(n_0,n_1,n_2\), respectively, with \(n=n_0+n_1+n_2\). Let $I_{\mathrm{train1}},$ $I_{\mathrm{train2}}$ and $I_{\mathrm{cal}}$ denote the corresponding index sets. These three subsets are used for prediction modeling, uncertainty estimation, and risk calibration, respectively, with separate training samples used
to avoid bias in residual-based uncertainty estimation. Using $\mathcal D_{\mathrm{train1}}$ and $\mathcal D_{\mathrm{train2}}$, respectively, we estimate $\mu(x)=\mathbb E(Y\mid X=x)$ by $\widehat\mu_{n_0}(x)$ and construct a positive local uncertainty estimate $\widehat\sigma_{n_1}(x)$. We define the \emph{UNEC score} as $$Z(x;c):=\frac{\hat\mu_{n_0}(x)-c}{\hat\sigma_{n_1}(x)}.$$
For $i\in I_{\mathrm{cal}}\cup I_{\mathrm{test}}$, the score can be calculated as $Z_i=Z(X_i;c)$. The normalization expresses the fitted margin in units of local predictive uncertainty, thus larger scores provide stronger standardized evidence of exceeding $c$. For a cutoff $z$, define
$
\widehat S(z)=\{j\in I_{\mathrm{test}}:Z(X_{n+j};c)\ge z\}.
$

For any candidate threshold $z\in\mathbb{R}$, define the empirical MDR and SDR on the calibration set as
$$
\widehat{\mathrm{MDR}}_{n_2}(z)=\frac{1}{n_2}\sum_{i\in I_{\mathrm{cal}}}L_{i,1}\mathbbm{1}\{Z_i\ge z\},\qquad
\widehat{\mathrm{SDR}}_{n_2}(z)=\frac{\sum_{i\in I_{\mathrm{cal}}}L_{i,1}\mathbbm{1}\{Z_i\ge z\}}{1\vee\sum_{i\in I_{\mathrm{cal}}}\mathbbm{1}\{Z_i\ge z\}}.
$$
For a risk criterion \(\mathcal E\in\{\mathrm{MDR},\mathrm{SDR}\}\), let \(\mathcal Z_{\mathrm{cand}}=\{Z_i:i\in I_{\mathrm{cal}}\}\) be the set of candidate cutoffs. Since the power of the thresholding rule \(\widehat{S}(z)\) is non-increasing in \(z\), the final cutoff is chosen as
$\widehat{Z}_{\mathcal E}=\min\{z\in \mathcal Z_{\mathrm{cand}}:\widehat{\mathcal{E}}_{n_2}(z)\le q\},$
if this set is nonempty, and \(\widehat{Z}_{\mathcal E}=\infty\) otherwise. The final selection set is then
$
\widehat{S}_{\mathcal E}=\{j \in I_{\mathrm{test}}:Z_j^{\mathrm{test}}\ge \widehat{Z}_{\mathcal E}\}.
$
The complete procedure is summarized in Algorithm~\ref{algo:UNEC_constant}.

\begin{algorithm}[ht]
    \caption{Uncertainty-Normalized Empirical Calibration (UNEC) selection procedure}
    \label{algo:UNEC_constant}
    \KwIn{Observed set $\mathcal{D}_{\mathrm{obs}}=\{X_i, Y_i\}_{i=1}^n$, test set $\mathcal{D}_{\mathrm{test}}=\{X_{n+j}\}_{j=1}^m$, sample sizes $n_0, n_1, n_2$ of $\mathcal{D}_{\mathrm{train1}}, \mathcal{D}_{\mathrm{train2}}$ and $\mathcal{D}_{\mathrm{cal}}$ with $n_0 + n_1 + n_2=n$, common threshold $c$, target risk level $q \in (0,1)$, risk criterion $\mathcal{E}\in\{\mathrm{MDR},\mathrm{SDR}\}$.}
    \KwOut{Selection set $\widehat{S}_{\mathcal E}$.}
    
    Randomly split $\mathcal{D}_{\mathrm{obs}}$ into three disjoint subsets $\mathcal{D}_{\mathrm{train1}}, \mathcal{D}_{\mathrm{train2}}$ and $\mathcal{D}_{\mathrm{cal}}$ of sizes $n_0, n_1$ and $n_2$, respectively.
    
    Fit a prediction model $\hat{\mu}_{n_0}(\cdot)$ using $\mathcal{D}_{\mathrm{train1}}$, and fit an uncertainty estimator $\hat{\sigma}_{n_1}(\cdot)$ using $\mathcal{D}_{\mathrm{train2}}$.
    
    For each $i\in I_{\mathrm{cal}}$, compute the uncertainty-normalized score $Z_i = \frac{\hat{\mu}_{n_0}(X_i)-c}{\hat{\sigma}_{n_1}(X_i)}$ and the corresponding loss $L_{i}=\ell(Y_{i},c)$, and set $\mathcal{Z}_{\mathrm{cand}} = \{Z_i:i\in I_{\mathrm{cal}}\}.$
    
    Set $\widehat{Z}_{\mathcal E} = \min \{z\in\mathcal{Z}_{\mathrm{cand}}: \widehat{\mathcal{E}}_{n_2}(z)\le q\}$, with $\widehat{Z}_{\mathcal E} = \infty$ if the feasible set is empty.
    
    For each $j \in I_{\mathrm{test}}$, compute the test score $Z_j^{\mathrm{test}}=\frac{\hat{\mu}_{n_0}(X_{n+j})-c}{\hat{\sigma}_{n_1}(X_{n+j})}.$
    
    Return the estimated discovery set $\widehat{S}_{\mathcal E} = \{j \in I_{\mathrm{test}}:Z_j^{\mathrm{test}}\ge\widehat{Z}_{\mathcal E}\}$.
    
\end{algorithm}

\paragraph{Observation-specific thresholds}
Observation-specific thresholds arise in two settings. When a target $C_i$ is observed for each calibration and test observation, UNEC directly calibrates the paired scores $Z_i=\{\widehat\mu(X_i)-C_i\}/\widehat\sigma(X_i)$ using the corresponding losses $\ell(Y_i,C_i)$. When only test-specific thresholds $c_j$ are available, the calibration sample is instead re-evaluated for each target $c_j$. Both extensions are described in Appendix~\ref{app:varying_thresholds}.

\paragraph{Conservative calibration} \label{para:conservative}
The preceding rule selects the least conservative cutoff satisfying the empirical MDR or SDR constraint. In finite samples, however, the empirical risk curve may underestimate its population counterpart. Thus we introduce a \(\delta\)-adjusted version of UNEC inspired by \citet{barber2015knockoffs} and \citet{qin2025revamping}. For simplicity, we describe the adjustment under the common-threshold setting.

Write $L_i=\ell(X_i,Y_i,c)$ and \(Z_i=Z(X_i;c)\) for \(i\in I_{\mathrm{cal}}\). Define the adjusted empirical MDR and SDR as
$$
\widehat{\mathrm{MDR}}_{n_2}(z,\delta):=\frac{1}{n_2}\sum_{i\in I_{\mathrm{cal}}}L_i\mathbbm{1}\{Z_i\ge z\}+\frac{\delta}{n_2},\qquad
\widehat{\mathrm{SDR}}_{n_2}(z,\delta):=\frac{\sum_{i\in I_{\mathrm{cal}}}L_i\mathbbm{1}\{Z_i\ge z\}+\delta}{1\vee\sum_{i\in I_{\mathrm{cal}}}\mathbbm{1}\{Z_i\ge z\}
}.
$$
The conservative procedure replaces $\widehat{\mathcal{E}}_{n_2}(z)$ in the preceding cutoff rule by $\widehat{\mathcal{E}}_{n_2}(z,\delta)$, with larger $\delta$ yielding more conservative selections. For SDR, the adjustment improves finite-sample stability while the formal guarantee remains asymptotic. For MDR, however, $\delta\ge 1-q$ yields finite-sample risk control by Theorem~\ref{thm:finite_mdr}.

\paragraph{Two-sided selection with interval targets}
\label{para:two_sided}
The UNEC framework can also be adapted to two-sided selection problems, where the scientific goal is to select observations whose responses fall within a target interval $(a,b)$ with $a<b$. For a common interval, the hypotheses are
$$
H_{0j}:Y_{n+j}\in(-\infty,a)\cup(b,+\infty),
\qquad
H_{1j}:Y_{n+j}\in[a,b],
\qquad j=1,\ldots,m.
$$
Unlike one-sided selection, interval targets favor predictions lying far from both boundaries
relative to local predictive uncertainty. To this end, we define the standardized safety-margin score
\begin{equation}
\label{eq:two_sided_score}
\widetilde Z(x;a,b)=\frac{\min\{\hat\mu_{n_0}(x)-a,\;b-\hat\mu_{n_0}(x)\}}{\hat\sigma_{n_1}(x)}.
\end{equation}
This score is positive when the fitted mean lies inside the interval and increases with the standardized distance from the nearer boundary. Using this score and the corresponding interval loss, the same empirical MDR/SDR calibration and threshold-selection steps in Algorithm~\ref{algo:UNEC_constant} apply. The risk guarantees and observation-specific interval extensions are presented in Appendix~\ref{app:varying_thresholds}.

\section{Theoretical Guarantees} \label{sec:theory}

\subsection{Risk control and power analysis}\label{subsec:asymptotic}
We establish asymptotic SDR control for UNEC with direct empirical calibration and finite-sample MDR control for its conservative variant. To simplify the discussion, we assume the common-threshold case with $c_j\equiv c$. Extensions of these risk-control results to observation-specific and interval targets are given in Appendix~\ref{app:varying_thresholds}.

Throughout this subsection, we condition on the training data, \(\mathcal D_{\mathrm{train1}}\) and \(\mathcal D_{\mathrm{train2}}\), and fitting randomness, so that $Z(\cdot;c)$ is fixed and the calibration and test observations remain i.i.d.. All probabilities, expectations, and convergence statements below are under this conditional law, and we suppress the conditioning in the notation for readability.

For an independent observation $(X,Y)$, write $Z=Z(X;c)$ and $L=\ell(X,Y,c)$. For a score cutoff $z\in\mathbb{R}$, define $F(z) = \mathbb{P}(Z \ge z),$ and $G(z) = \mathbb{E}\{L\cdot \mathbbm{1}\{Z \ge z\}\}.$
The population conditional risk at threshold $z$ is $Q(z) = \frac{G(z)}{F(z)}$ for $F(z) > 0$, and the population Power is $\text{Pow}(z) = \frac{\mathbb{P}(Z \ge z, Y \ge c)}{\mathbb{P}(Y \ge c)}$. 
Under the regularity conditions in Assumption~\ref{assump:unec_reg} (Appendix~\ref{app:asym_sdr_proof}), Theorem~\ref{thm:unec_asymptotic} establishes asymptotic SDR control and characterizes the limiting SDR and power.
\begin{colorboxyellow}
\begin{thm}[\emph{Asymptotic SDR Control}]\label{thm:unec_asymptotic}
Let $\widehat{Z}_{n_2}$ be the calibrated threshold obtained by UNEC at a target SDR level $q\in(0,1)$. Under Assumption~\ref{assump:unec_reg}, with $m$ fixed, as the calibration size $n_2\to\infty$, $\limsup_{n_2 \to \infty} \textnormal{SDR}_{\textnormal{test}}(\widehat{Z}_{n_2}) \le q$.
Furthermore,
$\widehat{Z}_{n_2}\xrightarrow{\mathrm{a.s.}}z^*$, where $z^*$ is defined in Assumption~\ref{assump:unec_reg}. The following limit identities hold, with the power statement requiring $\gamma>0$:
$$
\begin{aligned}
    &\lim_{n_2\to\infty} \mathrm{SDR}_{\mathrm{test}}(\widehat{Z}_{n_2})=q\left\{1-(1-\beta)^m\right\},\\
    &\lim_{n_2\to\infty} \mathrm{Power}_{\mathrm{test}}(\widehat{Z}_{n_2}) = \text{Pow}(z^*)\left\{1-\bigl(1-\gamma\bigr)^m\right\},
\end{aligned}
$$
where $\beta=\mathbb P\{Z\ge z^*\}=F(z^*)>0$ and $\gamma=\mathbb{P}(Y \ge c)$.

\end{thm}
\end{colorboxyellow}
We next establish finite-sample MDR control for the conservative version of UNEC.
\begin{colorboxyellow}
\begin{thm}[\emph{Finite-sample MDR control}]
\label{thm:finite_mdr}
Under the conditional law above, suppose that the calibration and test observations are exchangeable, and the loss $L$ takes values in $[0,1]$. If $\delta\ge 1-q$, then the conservative MDR procedure satisfies
$$
\mathbb E\left[L_{n+j,j}\mathbbm{1}\{Z(X_{n+j};c)\ge \widehat{Z}_{\mathrm{MDR},\delta}\}\right]\le q,\qquad j=1,\ldots,m.
$$
Consequently, $\mathrm{MDR}(\widehat{S}_{\mathrm{MDR},\delta})\le q.$

\end{thm}
\end{colorboxyellow}
The unadjusted MDR procedure ($\delta=0$) also achieves asymptotic MDR control at level $q$, as a consequence of Theorem~\ref{thm:finite_mdr}. An explicit finite-sample upper bound and its proof are provided in Corollary~\ref{cor:asymptotic_mdr} in Appendix~\ref{app:asymptotic_mdr}.

\subsection{Power optimality under the location--scale model}
\label{subsec:loc-scale}
For general bounded losses, the optimal ranking is generally loss-specific. We therefore establish power optimality for one-sided binary loss with a common threshold $c$, where $\ell(Y,c)=\mathbbm 1\{Y<c\}$ and SDR reduces to FDR. Assume the \emph{location--scale} model
$Y=\mu(X)+\sigma(X)\epsilon$,
where $\sigma(X)>0$, $\epsilon$ is independent of $X$, has a continuous cumulative distribution function $F_\epsilon$, and satisfies $\mathbb E(\epsilon)=0$ and $\operatorname{Var}(\epsilon)=1$.
Define the oracle UNEC score as $Z^o(X) := \frac{\mu(X)-c}{\sigma(X)}$. Under this model, the posterior success probability satisfies
$$
\eta(x) := \mathbb P(Y\ge c\mid X=x) = 1-F_\epsilon\{-Z^o(x)\}.
$$
Hence, $\eta(x)$ is nondecreasing in $Z^o(x)$, so the oracle UNEC score ranks observations by their posterior probability of exceeding the target threshold and provides the natural benchmark for power-efficient selection.
We next show that UNEC asymptotically attains the oracle power envelope under SDR/FDR control among observation-wise selection rules.
\begin{colorboxyellow}
\begin{thm}[\emph{Asymptotic power optimality}]
\label{thm:opt_power}
Suppose the location--scale model above holds and the regularity conditions in Appendix~\ref{app:power_optimality} are satisfied. Further assume $\sup_x|\hat\mu_{n_0}(x)-\mu(x)|\overset{p}\longrightarrow0, \sup_x|\hat\sigma_{n_1}(x)-\sigma(x)|\overset{p}\longrightarrow0, \sup_x|\mu(x)-c|<\infty,$ and $\inf_x\sigma(x)\ge \underline{\sigma}>0$. Then, UNEC asymptotically attains the oracle power envelope under SDR/FDR control among observation-wise selection rules.

\end{thm}
\end{colorboxyellow}
The result follows by showing that the oracle standardized margin induces an optimal ranking, characterizing the finite-$m$ SDR-constrained oracle rule, and establishing convergence of the estimated UNEC score and cutoff to their oracle counterparts. Detailed results and proofs are provided in Appendix~\ref{app:power_optimality}.

\section{Numerical Studies}\label{sec:simulation}

We evaluate the finite-sample risk control and selection power of UNEC under binary and general bounded losses. The experiments cover five data-generating processes (DGPs): three location--scale models and two covariate-dependent Gaussian mixture models that depart from this structure. Full experimental settings, DGP specifications, and details of the competing methods are provided in Appendices~\ref{app:simulation_setup} and~\ref{app:simulation_dgps}. Additional simulations with observation-specific thresholds and two-sided interval targets are reported in Appendices~\ref{app:simulation_varying_thresholds} and~\ref{app:simulation_two_sided}, respectively.

\paragraph{FDR control and power comparison}
We consider one-sided selection with a common threshold under binary loss, where SDR reduces to FDR. We compare UNEC with CS \citep{jin2023selection} using the clipped score, the NP method \citep{qin2025revamping} under a Gaussian location-shift working model, and UNEC-Raw. UNEC-Raw uses the same empirical calibration as UNEC but ranks candidates by the unnormalized margin $\widehat{\mu}(X)-c$.
\begin{figure}[t]
    \centering
    \includegraphics[width=0.95\linewidth]{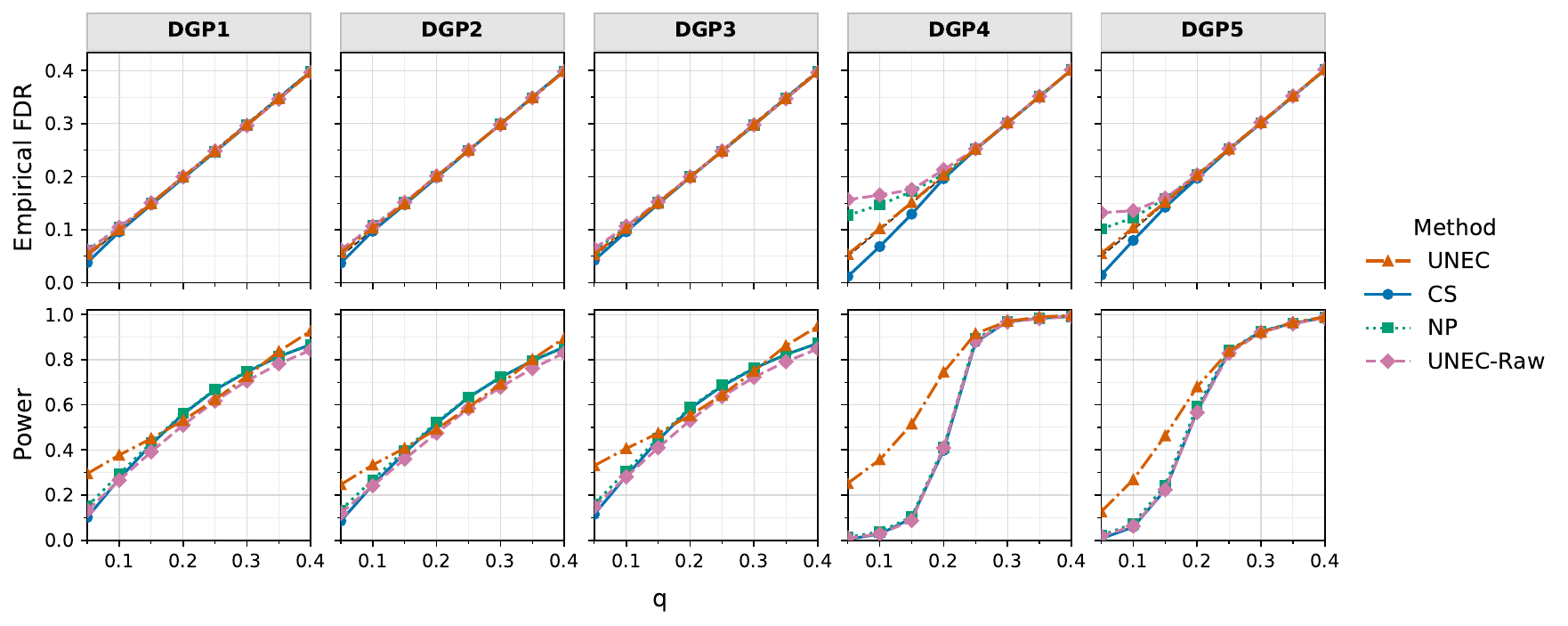}
    \caption{\textbf{FDR control and power across DGPs 1--5.} Empirical FDR (top row) and power (bottom row) are shown for UNEC, CS, NP, and UNEC-Raw as the nominal FDR level $q$ varies. The dashed black lines indicate the nominal FDR.}
    \label{fig:fdr_power}
\end{figure}

Figure~\ref{fig:fdr_power} shows that UNEC closely tracks the nominal FDR across all five DGPs over the tested range of $q$. CS and NP are similarly calibrated under DGPs 1--3, whereas under DGPs 4--5, CS becomes conservative while NP and UNEC-Raw can exceed the nominal level, particularly for small $q$. UNEC achieves competitive or higher power, with the most pronounced gains under DGPs 4--5 at stringent FDR levels, while these differences diminish as $q$ increases.

These gains reflect both empirical calibration, which allows UNEC to make fuller use of the available risk budget, and uncertainty normalization, which distinguishes candidates with similar fitted means but different predictive uncertainty. Appendix~\ref{app:fdr_comparison} further compares power against empirical FDR. UNEC retains favorable tradeoffs under DGPs 4--5, including an advantage over UNEC-Raw that supports the contribution from uncertainty-normalized ranking.

\paragraph{SDR control and power comparison}
We next compare the SDR control of UNEC and three reward-normalized SCoRE variants (SCoRE-dtm, SCoRE-homo, and SCoRE-hete) over all five DGPs \citep{bai2026csgeneralrisk}. We consider the sigmoid loss $\ell_\tau(y,c)=\{1+\exp[\tau(y-c)]\}^{-1}$, where $\tau=10$. 

\begin{figure}[t]
    \centering
    \includegraphics[width=0.95\linewidth]{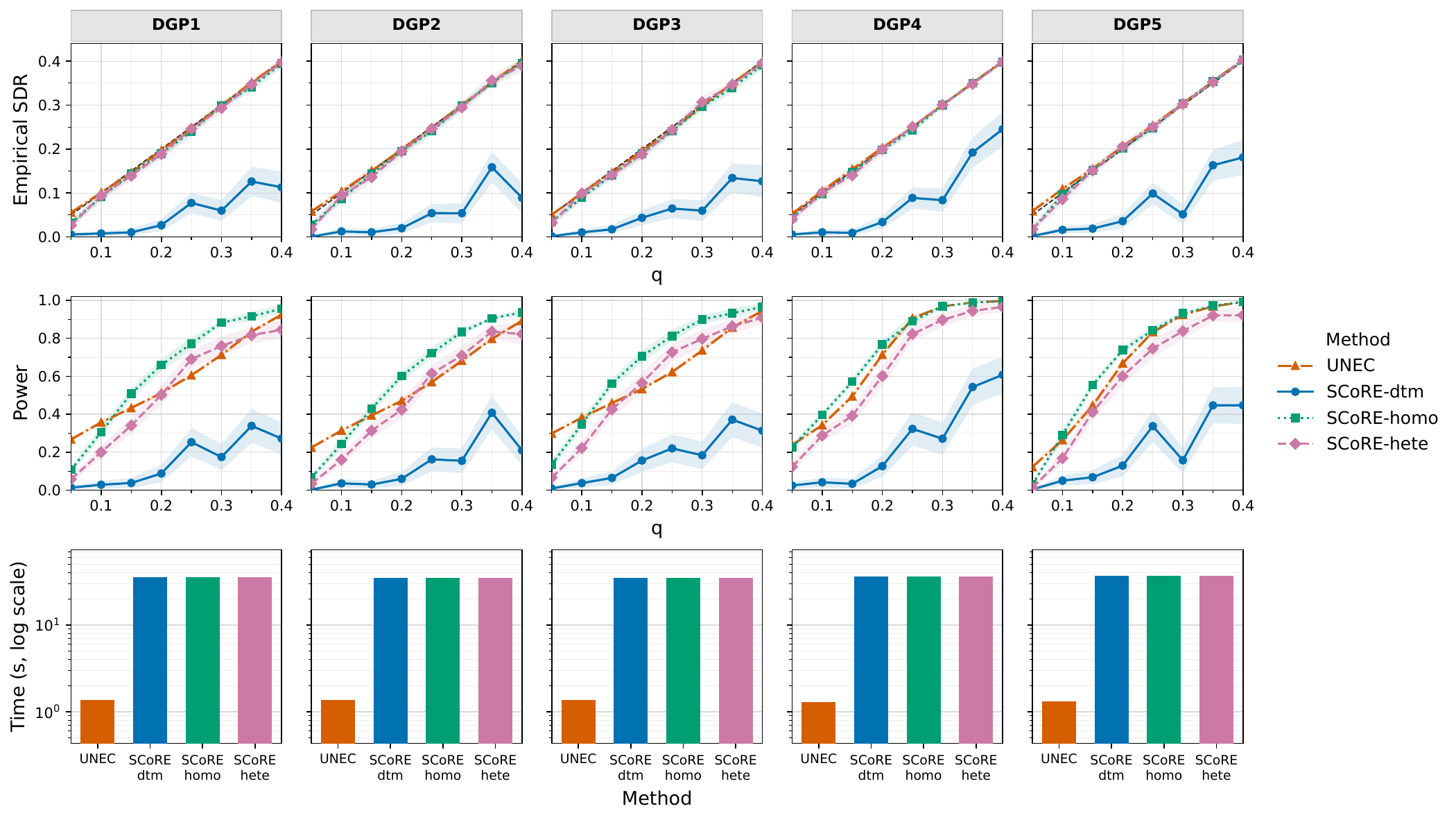}
    \caption{\textbf{SDR control, power, and computational efficiency across DGPs 1--5.} Empirical SDR (top row), power (middle row), and computation time (bottom row) are shown for UNEC and three SCoRE variants (SCoRE-dtm, SCoRE-homo, and SCoRE-hete). The nominal risk level $q$ varies in the SDR and power panels, with the dashed line indicating $y=q$. Computation time is reported in seconds on a logarithmic scale.}
    \label{fig:sdr_power_timing}
\end{figure}

Figure~\ref{fig:sdr_power_timing} compares UNEC with three SCoRE variants across five DGPs. UNEC maintains empirical SDR close to the nominal level while achieving competitive power. Its advantage is most pronounced at stringent risk levels. Under DGPs 1--3, UNEC has higher power for small $q$, while the gap narrows as $q$ increases. Under DGPs 4 and 5, UNEC remains broadly competitive with SCoRE-homo, whereas SCoRE-dtm is substantially more conservative and less powerful.

UNEC is also markedly more computationally efficient, requiring about 1.3 seconds per repetition over the full $q$ grid compared with 35--37 seconds for SCoRE. In our implementation, SCoRE is bottlenecked by the selection stage, which computes base e-values over candidate thresholds for each test observation, with dominant cost scaling approximately as $m(n+m)$. This gap becomes more important as sample sizes grow, giving UNEC a clear advantage for large-scale screening.

\section{Real Data Analysis for Drug Discovery}\label{sec:real_data}
We apply UNEC to three molecular-property screening tasks with different scientific targets: identifying highly soluble compounds, screening compounds under molecule-specific toxicity thresholds, and selecting compounds within a desirable lipophilicity interval. These applications cover a common one-sided threshold, observation-specific thresholds, and a two-sided target interval, and allow us to evaluate the method under both binary and general bounded losses.

\paragraph{Data and models.}
All datasets are obtained from the Therapeutics Data Commons (TDC) \citep{huang2021tdc}. Molecular structures are represented by SMILES strings and encoded using AttentiveFP through the DeepPurpose library \citep{huang2020deeppurpose,xiong2020attentivefp}. Details of data splitting, model training, cross-fitting, and evaluation protocols are provided in Appendix~\ref{app:implementation}.

\paragraph{Task 1: Screening for highly soluble compounds.}
Aqueous solubility is an important molecular property in drug development because insufficient solubility can limit absorption and bioavailability. We consider the AqSolDB dataset, which was constructed by integrating and curating experimental aqueous-solubility measurements from nine publicly available sources and contains \(9{,}982\) unique and valid compounds \citep{sorkun2019aqsoldb}. Let \(Y\) denote the measured aqueous solubility on the \(\mathrm{LogS}\) scale. Following the original AqSolDB study, we set $c=0$ and seek to identify compounds satisfying $Y>c$, corresponding to highly soluble compounds \citep{sorkun2019aqsoldb}. We use the binary false-selection loss $L=\mathbbm{1}\{Y\leq 0\},$ so that SDR coincides with FDR, and compare UNEC with CS and the NP method.

\begin{figure}[t]
    \centering
    \includegraphics[width=0.85\linewidth]{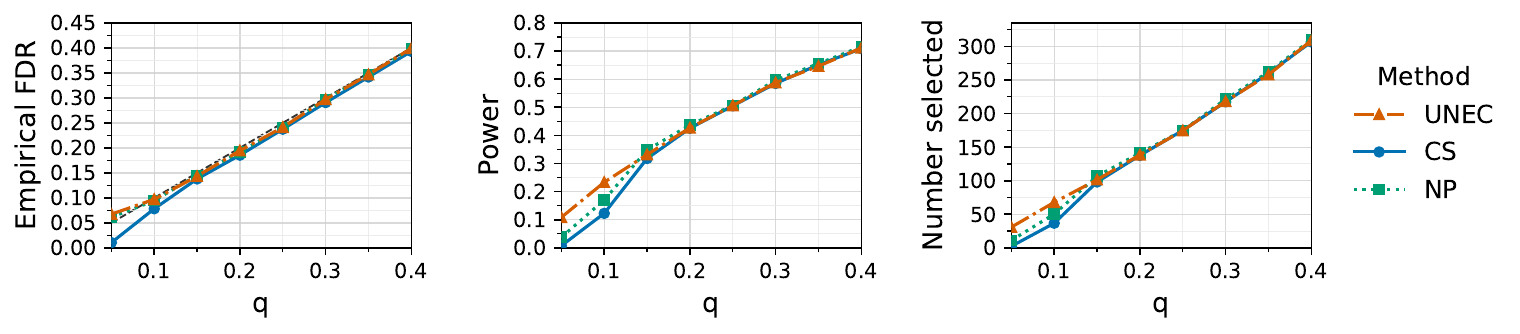}
    \caption{\textbf{FDR control and selection performance for aqueous-solubility screening.} Empirical FDR, power, and the average number of selected compounds are presented for CS, NP, and UNEC as the nominal FDR level \(q\) varies. The dashed black lines indicate $y=q$.}
    \label{fig:aqsoldb_fdr}
\end{figure}

Figure~\ref{fig:aqsoldb_fdr} shows that UNEC provides a highly competitive FDR--power trade-off, with its advantage most pronounced at stringent risk levels. Around $q=0.1$, UNEC closely tracks the nominal FDR while achieving substantially higher power and selecting more compounds than CS and NP. CS is more conservative in this regime, while the differences among the three methods gradually diminish as $q$ increases. These results indicate that accounting for molecule-specific predictive uncertainty improves the ranking of promising compounds and leads to more efficient screening, particularly when stringent FDR control is required.

\paragraph{Task 2: Toxicity screening with molecule-specific thresholds.}
Assessing acute toxicity is important for prioritizing compounds for further development. We next consider the LD50\_Zhu dataset, which contains experimental rat acute oral toxicity measurements for $7{,}385$ organic compounds \citep{zhu2009acute}. Let $D_j$ denote the median lethal dose ($\mathrm{LD}_{50}$) in mg/kg and $\mathrm{MW}_j$ the molecular weight in g/mol. Larger $\mathrm{LD}_{50}$ values indicate lower acute toxicity. Following the Globally Harmonized System of Classification and Labelling of Chemicals (GHS), we target compounds with $D_j>2000$ mg/kg, where $2000$ mg/kg is the upper boundary of acute oral toxicity Category~4 \citep{un2017ghs}.

TDC reports toxicity as the negative base-10 logarithm of $\mathrm{LD}_{50}$ measured in mol/kg. We reverse the sign and define $Y_j=\log_{10}\left(\frac{D_j}{1000\,\mathrm{MW}_j}\right),$ and $c_j=\log_{10}\left(\frac{2000}{1000\,\mathrm{MW}_j}\right)$,
so that $Y_j>c_j$ is equivalent to $D_j>2000$ mg/kg. Because the molecular weight, and hence the corresponding threshold $c_i$, is available for every calibration and test compound, we apply the paired observation-specific version of UNEC described in Appendix~\ref{app:varying_thresholds}. To allow graded penalties for compounds falling below the target, we use
$
\ell_\tau(Y_j,c_j) = \min\{1,\tau(c_j-Y_j)_+\},
$
where $\tau=10$. We compare UNEC with three variants of SCoRE (SCoRE-dtm, SCoRE-homo, and SCoRE-hete).

\begin{figure}[t]
    \centering
    \includegraphics[width=\linewidth]{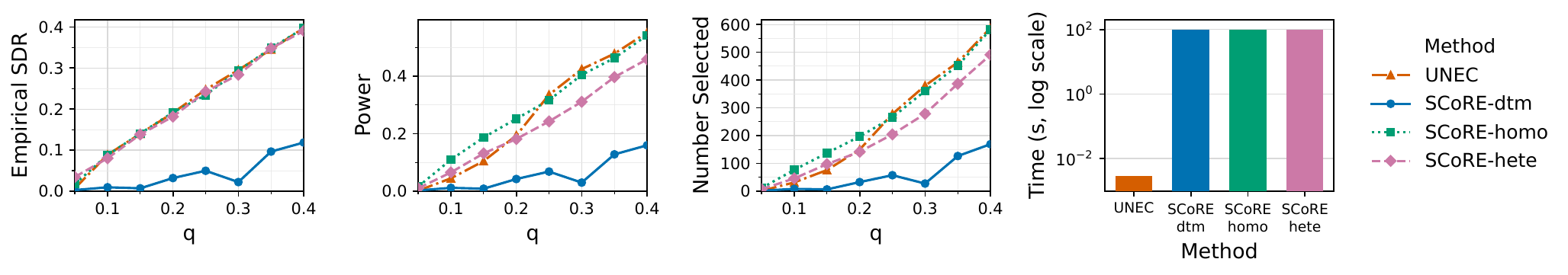}
    \caption{\textbf{SDR control, power, selection size, and computational efficiency for toxicity screening with molecule-specific thresholds.} Empirical SDR, power, average number of selected compounds, and computation time are shown for UNEC and three SCoRE variants (SCoRE-dtm, SCoRE-homo, and SCoRE-hete) as the nominal SDR level $q$ varies. The dashed line indicates $y=q$ in the SDR panel. Computation time is reported in seconds on a logarithmic scale.}
    \label{fig:ld50_varying_sdr}
\end{figure}

Figure~\ref{fig:ld50_varying_sdr} compares UNEC with three SCoRE variants for toxicity screening with molecule-specific thresholds. UNEC tracks the nominal SDR level closely across the considered values of $q$, as do SCoRE-homo and SCoRE-hete, whereas SCoRE-dtm is markedly more conservative. In terms of screening performance, UNEC remains competitive with SCoRE-homo in both power and the number of selected compounds, while substantially outperforming the more conservative SCoRE-dtm. SCoRE-hete generally selects fewer compounds and attains lower power than UNEC at moderate and large $q$.

The computational difference is much more pronounced. UNEC requires only a small fraction of the runtime of the SCoRE variants, whose cost is dominated by repeated e-value calculations during selection. Thus, in this molecule-specific screening task, UNEC achieves competitive risk--power performance while offering a substantial computational advantage, which is particularly relevant for large-scale compound screening.

\paragraph{Task 3: Screening for compounds within a desirable lipophilicity range.}
Lipophilicity influences both aqueous solubility and membrane permeability, making an intermediate range desirable for drug development. We finally consider the Lipophilicity\_AstraZeneca dataset, which contains experimental lipophilicity measurements for \(4{,}200\) compounds, and was subsequently included in the MoleculeNet benchmark \citep{wenlock2015dmpk,wu2018moleculenet}. Let \(Y\) denote the logarithm of the octanol--water distribution coefficient measured at pH \(7.4\), commonly written as \(\log D_{7.4}\). Guided by the favorable \(\log D\) range of approximately \(1\)--\(3\) discussed by \citet{waring2010lipophilicity}, we seek to identify compounds satisfying \(Y\in[1,3]\). 
To penalize departures from either boundary, we use the bounded distance loss
$
\ell_\tau(Y;[1,3])=\min\left\{1,\tau\left[(1-Y)_+ +(Y-3)_+\right]\right\} 
$, $\tau\in\{1,2,5\}$. We also consider the binary loss \(\ell_\infty(Y;[1,3])=\mathbbm{1}\{Y\notin[1,3]\}\), under which SDR coincides with FDR. We apply the two-sided UNEC procedure under SDR control with \(\delta=1\).

\begin{figure}[ht]
    \centering
    \includegraphics[width=0.85\linewidth]{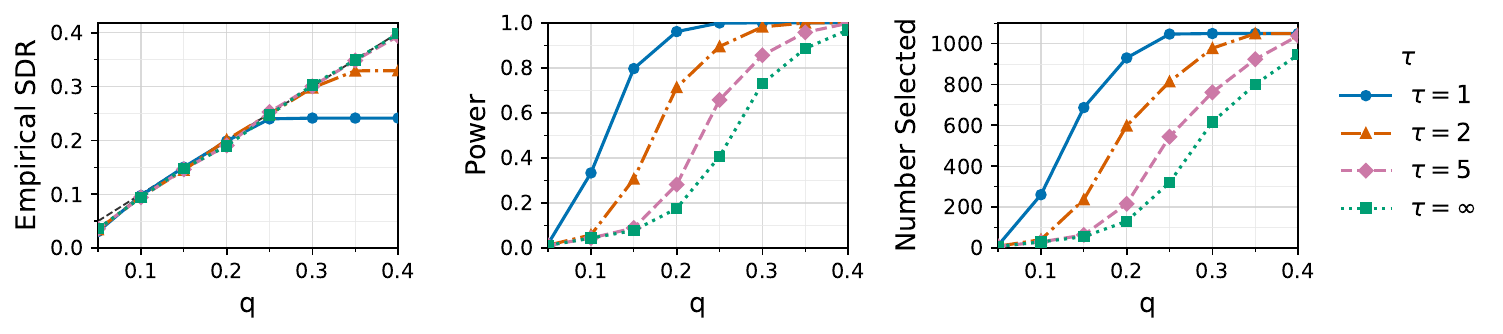}
    \caption{\textbf{SDR control and selection performance for two-sided lipophilicity screening.} Empirical SDR, power, and the average number of selected compounds are presented for UNEC as the nominal SDR level $q$ varies. The four curves correspond to $\tau=1,2,5,\infty$, where $\tau=\infty$ denotes binary loss. The dashed black line in the first panel indicates $y=q$.}
    \label{fig:lipophilicity_two_sided_sdr}
\end{figure}

Figure~\ref{fig:lipophilicity_two_sided_sdr} shows that empirical SDR generally stays close to or below the nominal level, and also highlights the effect of graded penalties in interval screening. Smaller $\tau$ yields higher power and more selections. As $q$ increases, the small-$\tau$ rules eventually select nearly all test compounds, so power approaches one, and SDR levels off near the average loss of the full test population. Larger $\tau$ penalizes departures from the interval more strongly and approaches the binary-loss setting. Corresponding MDR results are reported in Appendix~\ref{app:rda_mdr}.

\section{Discussion}
UNEC combines uncertainty-normalized ranking with direct empirical loss calibration. As shown in Section~\ref{subsec:loc-scale}, for a common one-sided target, under the location--scale model the oracle UNEC score preserves the ordering of the posterior success probability $\eta(x)=\mathbb P(Y\ge c\mid X=x)$. The importance of this posterior success probability has also been recognized in conformal selection. \citet{jin2023selection} recommend scores that preserve its ordering, including direct probability estimation.

A similar connection holds with the NP method of \citet{qin2025revamping}. Under one-sided binary loss and a location--scale model,
$
R_{\mathrm{NP}}(x;c)\propto\frac{1-\eta(x)}{\eta(x)},
$
thus smaller NP scores and larger oracle UNEC scores favor the same posterior ordering, but through different constructions. UNEC derives it from conditional location and scale without explicitly modeling the common residual distribution.

SCoRE \citep{bai2026csgeneralrisk} provides another related perspective. Its asymptotic SDR analysis identifies the ordering criterion $(l(x)-q)/r(x)$, where $l(x)$ and $r(x)$ are the conditional expected loss and reward, respectively, and smaller values are preferred. Under binary loss and reward $\mathbbm 1\{Y\ge c\}$,
$
\frac{l(x)-q}{r(x)} = \frac{1-q}{\eta(x)}-1,
$
which decreases in $\eta(x)$. Hence, in this setting, the SCoRE criterion and the oracle UNEC score agree on the preference for candidates with larger posterior success probabilities.

These shared ordering principles do not imply identical selections, because the calibration mechanisms differ. CS constructs conformal $p$-values followed by multiple testing, SCoRE uses risk-adjusted conformal $e$-values, whereas UNEC calibrates the selection cutoff directly from empirical losses.

The current theory has two main limitations. SDR control is asymptotic, and the power-optimality
result is restricted to one-sided binary loss under a location--scale model. Extending finite-sample selective-risk control and power characterization to more general losses and models is an important direction for future work.

\section*{Ethics Statement}

This work adheres to the ICLR Code of Ethics. This work develops statistical methodology for risk-controlled selection using simulated data and existing publicly available molecular-property datasets, including AqSolDB, LD50\_Zhu, and Lipophilicity\_AstraZeneca. No personal data were collected, and no new experiments involving human participants or animals were conducted. The molecular screening applications illustrate candidate prioritization and do not establish the safety or efficacy of individual compounds. Downstream use requires appropriate experimental validation, and the risk guarantees apply under the assumptions stated in the paper.

\section*{Reproducibility Statement}

The UNEC algorithm is described in Section~\ref{sec:method}, with theoretical assumptions and guarantees in Section~\ref{sec:theory} and proofs in Appendix~\ref{app:proofs}. Section~\ref{sec:simulation} presents the simulation studies, with detailed experimental protocols, competing methods, and data-generating processes in Appendices~\ref{app:simulation_setup} and~\ref{app:simulation_dgps}. Section~\ref{sec:real_data} identifies the datasets and screening targets, while Appendix~\ref{app:implementation} details data splitting, preprocessing, model training, and evaluation procedures. The accompanying code is available in an anonymous repository at \url{https://anonymous.4open.science/r/review-code-package-3620}.

\section*{AI Use Statement}

The authors independently conceived and developed the research questions, original ideas, scientific hypotheses, and core methodological and conceptual framework of UNEC. Generative AI was not used to identify the original research topic or research gap, brainstorm the core research ideas, or propose the paper title or keywords.

Generative AI tools were used to assist with method implementation, code writing and debugging, simulation-data generation code, data cleaning and reformatting, translation, and interpretation of results. Simulation data were generated programmatically from author-specified data-generating processes. AI assistance also included feedback on research methodology and experiments, discussion of experimental settings and parameters, refinement of mathematical statements and their assumptions, assistance with intermediate proof arguments, and drafting, checking, and revising proof text.

Additional uses included searching for and identifying relevant literature, summarizing and comparing existing work, drafting sections of the manuscript, improving language and readability, suggesting manuscript organization, formatting references, modifying scientific figures through plotting code, and preparing research artifacts such as scripts, tables, and reproducibility documentation. Survey or interview question design, transcription of research recordings, and qualitative or thematic data analysis were not applicable to this work.

All AI-assisted content, including text, mathematical statements and proofs, code, data-processing procedures, figures, interpretations of results, and references, was manually reviewed and checked by the authors. The authors take full responsibility for the accuracy, originality, and integrity of the final manuscript and accompanying research artifacts.

\bibliographystyle{iclr2027_conference}
\bibliography{Bibliography}

\appendix

\section{Extensions to Observation-Specific Thresholds}\label{app:varying_thresholds}

A more general setting arises when the hypothesis thresholds are not constant but vary across observations, that is, $H_{0j}: Y_{n+j} < c_j, H_{1j}: Y_{n+j} \ge c_j, j=1,\cdots,m.$ Here \(c_j\) may be fixed in advance or randomly generated, as long as it is observed before the selection decision is made. When thresholds vary across hypotheses, the notion of “exceeding the threshold” becomes target-dependent, and thus the score must also depend on the specific test index $j$.

Observation-specific thresholds arise in two settings: targets paired with every calibration and test observation, and external targets specified only for the test observations. We describe these settings separately below.

\subsection{Paired observation-specific thresholds}
\label{app:paired_thresholds}

The common-threshold procedure can be extended directly to targets paired with individual observations. For each calibration and test observation, let $C_i$ be an observed target threshold. Conditional on the training data and fitting randomness, assume that the triples $(X_i,Y_i,C_i)$ are i.i.d. The threshold may depend on the covariates, for example $C_i=g(X_i)$, while the function $g$ itself need not be known as long as the realized thresholds are observed for both calibration and test observations.

For each calibration observation, use the paired score 
$$
Z_i=Z(X_i;C_i)=\frac{\widehat\mu_{n_0}(X_i)-C_i}{\widehat\sigma_{n_1}(X_i)}
$$
and loss $L_i=\ell(X_i,Y_i,C_i)$. The empirical MDR and SDR are then computed exactly as in the common-threshold procedure using the paired values $(Z_i,L_i)$, yielding a single calibrated cutoff $\widehat Z_{\mathcal E}$. For each test observation $j \in I_{\mathrm{test}}$,
$$
Z_j^{\mathrm{test}} = \frac{\widehat\mu_{n_0}(X_{n+j})-C_{n+j}}{\widehat\sigma_{n_1}(X_{n+j})},
$$
and we finally select
$$
\widehat S_{\mathcal E}=\{j:Z_j^{\mathrm{test}}\ge\widehat Z_{\mathcal E}\}.
$$

\begin{remark}[Reduction of paired targets to a common threshold]
\label{rem:paired_reduction}
Suppose that the calibration and test triples $(X_i,Y_i,C_i)$ are i.i.d. copies of $(X,Y,C)$, independent of the training data and fitting randomness. Define
$$
\widetilde X_i=(X_i,C_i),\qquad \widetilde Y_i=Y_i-C_i.
$$
Then $Y_i\ge C_i$ is equivalent to $\widetilde Y_i\ge0$. The fitted location and scale functions $\widehat{\widetilde\mu}(x,c)=\widehat\mu(x)-c$ and $\widehat{\widetilde\sigma}(x,c)=\widehat\sigma(x)$ yield
$$
\frac{\widehat{\widetilde\mu}(X_i,C_i)}{\widehat{\widetilde\sigma}(X_i,C_i)} = \frac{\widehat\mu(X_i)-C_i}{\widehat\sigma(X_i)}.
$$
For a bounded loss $\ell(X,Y,C)$, define $\widetilde\ell((x,c),y)=\ell(x,y+c,c)$. Thus, the paired procedure is the common-threshold procedure at threshold zero for the transformed observations, with identical selection decisions, losses, and target indicators.

Consequently, the finite-sample MDR guarantee in Theorem~\ref{thm:finite_mdr} and the unadjusted MDR bound in Corollary~\ref{cor:asymptotic_mdr} apply with the same adjustment
requirements. Under Assumption~\ref{assump:unec_reg} for the conditional distribution of the paired score and loss, Theorem~\ref{thm:unec_asymptotic} also applies, with its power formula using $\gamma=\mathbb P(Y\ge C)>0$ and $Pow(z)=\mathbb P\{Z(X;C)\ge z\mid Y\ge C\}$. These risk guarantees average over the paired population and do not assert separate control conditional on each realized target value.

For binary loss $L=\mathbbm{1}\{Y<C\}$, additionally assume the location--scale model of Section~\ref{subsec:loc-scale} with $\epsilon$ independent of $(X,C)$. The transformed response then satisfies
$$
\widetilde Y = \widetilde\mu(\widetilde X)+\widetilde\sigma(\widetilde X)\epsilon,
$$
where $\widetilde\mu(x,c)=\mu(x)-c,$ and $\widetilde\sigma(x,c)=\sigma(x).$ Hence, under the asymptotic regime and assumptions of Theorem~\ref{thm:opt_power} applied to this transformed problem, its power-optimality conclusion extends to measurable observation-wise rules $S(X,C)$. In particular, the oracle regularity conditions concern $Z^o(X,C)=\{\mu(X)-C\}/\sigma(X)$, and the boundedness condition becomes
$$
\sup_{(x,c)\in\operatorname{supp}(X,C)}|\mu(x)-c|<\infty.
$$
The stronger independence condition holds automatically when $C=g(X)$ and $\epsilon$ is independent of $X$.
\end{remark}

\subsection{One-sided selection with test-specific external thresholds}
\label{app:test_specific_thresholds}

We next consider the setting in which each test observation has a specified target $c_j$, but no corresponding observation-specific threshold is available for the calibration observations. In this case, the calibration sample cannot be directly paired with observation-specific targets. Instead, for each test threshold $c_j$, we re-evaluate the full calibration sample relative to that target.

Using the same fitted functions \(\hat\mu_{n_0}(\cdot)\) and \(\hat\sigma_{n_1}(\cdot)\) from Section~\ref{sec:method}, define
$$
Z_{i,j} := Z\left(X_i,c_j\right) = \frac{\hat{\mu}_{n_0}(X_i)-c_j}{\hat{\sigma}_{n_1}(X_i)}, \quad L_{i,j}=\ell(Y_i,c_j) \qquad i\in I_{\mathrm{cal}}, \quad j\in I_{\mathrm{test}},
$$
Thus, for each test hypothesis $j$, the calibration points are re-evaluated relative to its own threshold \(c_j\). The corresponding candidate set is $\mathcal Z_{\mathrm{cand},j}=\{Z_{i,j}:i\in I_{\mathrm{cal}}\}.$ Meanwhile, target-specific empirical MDR and SDR become
$$
\begin{aligned}
    \widehat{\mathrm{MDR}}_{n_2,j}(z)&=\frac{1}{n_2}\sum_{i\in I_{\mathrm{cal}}}L_{i,j}\mathbbm{1}\{Z_{i,j}\ge z\},\\
    \widehat{\mathrm{SDR}}_{n_2,j}(z)&=\frac{\sum_{i\in I_{\mathrm{cal}}}L_{i,j}\mathbbm{1}\{Z_{i,j}\ge z\}}{1\vee\sum_{i\in I_{\mathrm{cal}}}\mathbbm{1}\{Z_{i,j}\ge z\}}.
\end{aligned}
$$
As described in Section~\ref{sec:method}, we also choose the smallest feasible cutoff \(\widehat{Z}_{\mathcal E,n_2,j}=\inf\{z\in\mathcal Z_{\mathrm{cand},j}:\) \(\widehat{\mathcal{E}}_{n_2,j}(z)\le q\},\) if the set is nonempty, and \(\widehat{Z}_{\mathcal E,n_2,j}=\infty\) otherwise. Finally, we select \(\widehat{S}_{\mathcal E}=\{j:Z_j^{\mathrm{test}}=\) \(\frac{\hat\mu_{n_0}(X_{n+j})-c_j}{\hat\sigma_{n_1}(X_{n+j})}\ge \widehat{Z}_{\mathcal E,n_2,j}\}.\) 
The detailed procedure is summarized in Algorithm~\ref{algo:UNEC_varying}. Intuitively, Algorithm~\ref{algo:UNEC_varying} can be viewed as applying the common-threshold UNEC procedure separately for each test-specific threshold \(c_j\), and then combining the resulting decisions into the final selection set. For each \(j\), the full calibration set \(\mathcal D_{\mathrm{cal}}\) is reused to construct the target-specific empirical risk curve and cutoff. This provides a simple and natural extension of Algorithm~\ref{algo:UNEC_constant} to observation-specific thresholds. Section~\ref{subsec:asymptotic} establishes the corresponding asymptotic risk-control guarantee and power characterization.
\begin{algorithm}[ht]
    \caption{UNEC selection procedure with observation-specific one-sided thresholds}
    \label{algo:UNEC_varying}
    \KwIn{Observed set $\mathcal{D}_{\mathrm{obs}}=\{X_i, Y_i\}_{i=1}^n$, test set $\mathcal{D}_{\mathrm{test}}=\{X_{n+j}\}_{j=1}^m$, sample sizes $n_0, n_1, n_2$ of $\mathcal{D}_{\mathrm{train1}}, \mathcal{D}_{\mathrm{train2}}$ and $\mathcal{D}_{\mathrm{cal}}$ with $n_0 + n_1 + n_2=n$, observation-specific thresholds $\{c_j\}_{j=1}^m$, target risk level $q \in (0,1)$, risk criterion $\mathcal{E}\in\{\mathrm{MDR},\mathrm{SDR}\}$.}
    \KwOut{Estimated discovery set $\widehat{S}_{\mathcal E}$.}
    
    Randomly split $\mathcal{D}_{\mathrm{obs}}$ into three disjoint subsets $\mathcal{D}_{\mathrm{train1}}, \mathcal{D}_{\mathrm{train2}}$ and $\mathcal{D}_{\mathrm{cal}}$ of sizes $n_0, n_1$ and $n_2$, respectively. 
    
    Fit a prediction model $\hat{\mu}(\cdot)$ using $\mathcal{D}_{\mathrm{train1}}$, and fit an uncertainty estimator $\hat{\sigma}(\cdot)$ using $\mathcal{D}_{\mathrm{train2}}$.
    
    Initialize \(\widehat{S}_{\mathcal E}=\emptyset\).\;

    \For{\(j=1,\ldots,m\)}{
    For each $i\in I_{\mathrm{cal}}$, compute the uncertainty-normalized calibration score $Z_{i,j} = \frac{\hat{\mu}_{n_0}(X_i)-c_j}{\hat{\sigma}_{n_1}(X_i)},$ and the corresponding loss $L_{i,j} = \ell(Y_{i},c_j).$ Let $\mathcal{Z}_{\mathrm{cand},j} = \left\{ Z_{i,j}:i\in I_{\mathrm{cal}}\right\}.$
    
    Set
    \(\widehat{Z}_{\mathcal E,j}=\min\left\{z\in\mathcal Z_{\mathrm{cand},j}:\widehat{\mathcal{E}}_{n_2,j}(z)\le q \right\},\) with \(\widehat{Z}_{\mathcal E,j}=\infty\) if the feasible set is empty.\;
    
    Compute the test score \(Z_j^{\mathrm{test}}=\frac{\hat\mu_{n_0}(X_{n+j})-c_j}{\hat\sigma_{n_1}(X_{n+j})}.\) \If {\(Z_j^{\mathrm{test}}\ge \widehat{Z}_{\mathcal E,j}\)}{Update \(\widehat{S}_{\mathcal E} \leftarrow \widehat{S}_{\mathcal E}\cup\{j\}.\)}
}
\Return{\(\widehat{S}_{\mathcal E}\).}
\end{algorithm}

\subsection{Theoretical guarantees for test-specific external thresholds}
\label{app:external_threshold_guarantees}

We study the external-threshold procedure in Appendix~\ref{app:test_specific_thresholds}. Let $\mathbf c=(c_1,\ldots,c_m)$ be fixed or, conditional on  the training data and fitting randomness, independent of all calibration and test observations. Throughout this subsection, we condition on the training data, fitting randomness, and $\mathbf c$. Thus, the fitted score functions and target thresholds are fixed. Unless explicitly stated otherwise, all probabilities, expectations, risks, and convergence statements below refer to this conditional law. Assume that the fitted UNEC scores are real-valued and that $L_{i,j}=\ell(Y_i,c_j)\in[0,1]$.

For $\delta\ge0$, define the target-specific adjusted empirical risks by
$$
\begin{aligned}
\widehat{\mathrm{MDR}}_{n_2,j}(z,\delta)&=\frac{1}{n_2}\left\{\sum_{i\in I_{\mathrm{cal}}} L_{i,j}\mathbbm{1}\{Z_{i,j}\ge z\}+\delta\right\},\\
\widehat{\mathrm{SDR}}_{n_2,j}(z,\delta)&=\frac{\sum_{i\in I_{\mathrm{cal}}}L_{i,j}\mathbbm{1}\{Z_{i,j}\ge z\}+\delta}{1\vee\sum_{i\in I_{\mathrm{cal}}}\mathbbm{1}\{Z_{i,j}\ge z\}}.
\end{aligned}
$$
For $\mathcal E\in\{\mathrm{MDR},\mathrm{SDR}\}$, let
$$
\widehat Z_{\mathcal E,\delta,j}=\min\left\{z\in\mathcal Z_{\mathrm{cand},j}:\widehat{\mathcal E}_{n_2,j}(z,\delta)\le q\right\},
$$
with $\widehat Z_{\mathcal E,\delta,j}=+\infty$ when the feasible set is empty. The corresponding selection set is
$$
\widehat S_{\mathcal E,\delta}=\left\{j:Z(X_{n+j};c_j)\ge\widehat Z_{\mathcal E,\delta,j}\right\}.
$$
The choice $\delta=0$ recovers Algorithm~\ref{algo:UNEC_varying}.

\paragraph{MDR control}
The finite-sample MDR guarantee extends directly to test-specific external thresholds. Suppose that the calibration and test observations are exchangeable under the conditional law specified above. Applying Theorem~\ref{thm:finite_mdr} separately to each fixed target $c_j$ and averaging the resulting expected-loss bounds over $j=1,\ldots,m$ yields, for every $n_2\ge 1$ and $m\ge 1$,
$$
\mathrm{MDR} \bigl(\widehat S_{\mathrm{MDR},\delta}\bigr) \le q,
\qquad \delta\ge1-q.
$$
Similarly, applying Corollary~\ref{cor:asymptotic_mdr} to each target with a single test observation and averaging gives
$$
\mathrm{MDR}\bigl(\widehat S_{\mathrm{MDR},0}\bigr) \le q+\frac{1-q}{n_2+1}.
$$
Thus, the unadjusted external-threshold procedure also achieves asymptotic MDR control at level $q$ as $n_2\to\infty$.

\paragraph{SDR control}
For the SDR analysis, let $(X,Y)$ be an independent observation from the common conditional distribution of the calibration and test observations. For each target $c_j$, write $Z_j=Z(X;c_j)$ and $L_j=\ell(Y,c_j)$, and define
$$
F_j(z)=\mathbb P(Z_j\ge z),
\qquad
G_j(z)=\mathbb E\left[L_j\mathbbm{1}\{Z_j\ge z\}\right].
$$
The population conditional risk is
$$
Q_j(z)=\frac{G_j(z)}{F_j(z)}, \qquad F_j(z)>0.
$$
Define the first population-feasible cutoff by
$$
z_j^*=\inf\left\{z\in\mathbb R:F_j(z)>0,\ Q_j(z)\le q\right\}.
$$
\begin{colorboxyellow}
\begin{thm}[\emph{Asymptotic SDR control for external thresholds}]
\label{thm:external_sdr}
Suppose that, under the conditional law specified above, the calibration and test observations are i.i.d. For every $j=1,\ldots,m$, suppose that Assumption~\ref{assump:unec_reg} holds with $(Z,L,F,G,Q,z^*)$ replaced by $(Z_j,L_j,F_j,G_j,Q_j,z_j^*)$. Under this conditional law, with the training stage, $\mathbf c$, and $m$ fixed, the following statements hold as the calibration size $n_2\to\infty$:
$$
\limsup_{n_2\to\infty}\mathrm{SDR}\bigl(\widehat S_{\mathrm{SDR}}\bigr) \le q.
$$
Furthermore,
$$
\widehat Z_{\mathrm{SDR},j}\xrightarrow{\mathrm{a.s.}}z_j^*,\qquad j=1,\ldots,m,
$$
and the following limit identity holds:
$$
\lim_{n_2\to\infty}\mathrm{SDR}\bigl(\widehat S_{\mathrm{SDR}}\bigr)=q\left\{1-\prod_{j=1}^m(1-\beta_j)\right\},
$$
where $\beta_j=F_j(z_j^*)>0$.
\end{thm}
\end{colorboxyellow}
The proofs of Theorems~\ref{thm:external_sdr} are provided in Appendix~\ref{app:external_threshold_proofs}.

\paragraph{Power optimality}
We consider binary loss $L_{i,j}=\mathbbm{1}\{Y_i<c_j\}$ under the location--scale model of Section~\ref{subsec:loc-scale}. For this power-optimality analysis, the target vector $\mathbf c$ and test size $m$ are fixed, while power is averaged over the training, calibration, and test data. For each target $j\in \{1,\cdots,m\}$, define the oracle score $Z_j^o(x)=\frac{\mu(x)-c_j}{\sigma(x)}.$ Under the location--scale model,
$$
\mathbb P(Y\ge c_j\mid X=x) = 1-F_\epsilon\{-Z_j^o(x)\}.
$$
Thus, Lemma~\ref{lem:oracle_ranking} applies separately to each target $c_j$.

To specify the comparison class, let $\mathbf s=(s_1,\ldots,s_m)$ consist of deterministic measurable rules $s_j:\mathcal X\to\{0,1\}$, and write $S(\mathbf s)=\{j:s_j(X_{n+j})=1\}.$ Define
$$
\mathcal A_q = \left\{\mathbf s: \mathbb E\left[\mathbbm{1}\{Y<c_j\}s_j(X)\right] \le q\,\mathbb E[s_j(X)], \quad j=1,\ldots,m\right\}.
$$
Whenever $\mathbb P\{s_j(X)=1\}>0$, the corresponding constraint is equivalent to $\mathbb P\{Y<c_j\mid s_j(X)=1\}\le q$. Rules that never select for a given target are also allowed.

For $F_j^o(z) := \mathbb P\{Z_j^o(X)\ge z\}>0$, define
$$
Q_j^o(z) = \mathbb P\{Y<c_j\mid Z_j^o(X)\ge z\},
$$
and let
$$
z_j^o = \inf\left\{z\in\mathbb R:\mathbb F_j^o(z)>0, Q_j^o(z)\le q\right\}.
$$
The corresponding oracle selection set is
$$
\widehat S^o=\left\{j:Z_j^o(X_{n+j})\ge z_j^o\right\}.
$$
\begin{colorboxyellow}
\begin{thm}[\emph{Power optimality under target-specific risk constraints}]
\label{thm:external_power_optimality}
Suppose that the training, calibration, and test observations are i.i.d. and satisfy the location--scale model of Section~\ref{subsec:loc-scale}. For every $j=1,\ldots,m$, assume that Assumption~\ref{assump:opt_power} holds for the common-target problem with $c=c_j$.

Suppose further that, as $n_0,n_1\to\infty$, $\sup_x |\widehat\mu_{n_0}(x)-\mu(x)|\xrightarrow{p}0, \sup_x |\widehat\sigma_{n_1}(x)-\sigma(x)| \xrightarrow{p}0,$ and that $\inf_x\sigma(x)\ge \underline \sigma>0, \max_{1\le j\le m} \sup_x|\mu(x)-c_j|<\infty.$ Then, as $n_0,n_1,n_2\to\infty$ while the test size $m$ and external target vector $\mathbf c$ remain fixed, the unadjusted SDR-calibrated external-threshold UNEC procedure satisfies
$$
\widehat Z_{\mathrm{SDR},j}\xrightarrow{p}z_j^o,\qquad j=1,\ldots,m,
$$
and
$$
\lim_{n_0,n_1,n_2\to\infty} \mathrm{Power} \bigl(\widehat S_{\mathrm{SDR}}\bigr) = \mathrm{Power}(\widehat S^o) = \sup_{\mathbf s\in\mathcal A_q}\mathrm{Power}\{S(\mathbf s)\}.
$$

\end{thm}
\end{colorboxyellow}
The optimality statement concerns observation-wise rules satisfying the target-specific constraints defining $\mathcal A_q$. These constraints imply overall SDR control, but are stronger than the single constraint $\mathrm{SDR}\le q$.

\subsection{Two-sided selection with observation-specific thresholds}

We now extend the two-sided procedure to observation-specific interval targets $T_j=(a_j,b_j)$ with $a_j<b_j$. Using the safety-margin score in \eqref{eq:two_sided_score}, the $j$-th test score is
$$
\widetilde Z_j^{\mathrm{test}}=\widetilde Z(X_{n+j};a_j,b_j).
$$
As in the one-sided setting, the calibration procedure depends on whether interval targets are also available for the calibration observations. When they are available, we calibrate a common cutoff using the paired scores and losses. Otherwise, we re-evaluate the calibration sample relative to each test interval and obtain a separate cutoff for each target.

\paragraph{Common or paired interval targets}
When a common interval $(a,b)$ is specified, or an interval target is observed for every calibration and test observation, we calibrate a single cutoff using the corresponding paired scores and losses. Let $T_i^{\mathrm{cal}}=(a_i^{\mathrm{cal}},b_i^{\mathrm{cal}})$ denote the interval associated with calibration observation $i$. For each $i\in I_{\mathrm{cal}}$, compute
$$
\widetilde Z_i=\widetilde Z(X_i;a_i^{\mathrm{cal}},b_i^{\mathrm{cal}}),
\qquad
L_i=\ell\{Y_i,T_i^{\mathrm{cal}}\}.
$$
Using these paired scores and bounded losses, we apply the same empirical MDR/SDR calibration and threshold-selection steps as in Algorithm~\ref{algo:UNEC_constant}, with candidate set $\widetilde{\mathcal Z}_{\mathrm{cand}}=\{\widetilde Z_i:i\in I_{\mathrm{cal}}\}$. This yields a common calibrated cutoff $\widehat Z_{\mathcal E}$ and the selection set
$$
\widehat S_{\mathcal E}=\{j:\widetilde Z_j^{\mathrm{test}}\ge\widehat Z_{\mathcal E}\},
\qquad
\mathcal E\in\{\mathrm{MDR},\mathrm{SDR}\}.
$$
The common-interval setting is recovered by taking every calibration and test interval to be $(a,b)$.

\paragraph{Test-specific external interval targets}
When interval targets are specified only for the test observations, we instead re-evaluate the full calibration sample relative to each test interval. For each target $T_j=(a_j,b_j)$, define
$$
\widetilde Z_{i,j}=\widetilde Z(X_i;a_j,b_j),
\qquad L_{i,j}=\ell\{Y_i,(a_j,b_j)\},
\qquad i\in I_{\mathrm{cal}}.
$$
We then apply the target-specific calibration steps in Algorithm~\ref{algo:UNEC_varying}, using $\widetilde{\mathcal Z}_{\mathrm{cand},j}=\{\widetilde Z_{i,j}:i\in I_{\mathrm{cal}}\}$, to obtain a separate cutoff $\widehat Z_{\mathcal E,j}$ for each target. The resulting selection set is
$$
\widehat S_{\mathcal E}=\{j:\widetilde Z_j^{\mathrm{test}}\ge\widehat Z_{\mathcal E,j}\}.
$$
Thus, the two-sided procedure follows the same calibration schemes as its one-sided counterpart, with the score and loss evaluated relative to the corresponding interval target.

The two-sided procedure provides a practical extension of UNEC to interval-target screening. We establish finite-sample MDR control for the adjusted procedure and asymptotic SDR control under the corresponding regularity conditions, and evaluate its selection power empirically.

\paragraph{Risk control}

The MDR and SDR guarantees extend to interval targets because the calibration arguments depend on the score and loss through their joint distribution, rather than the specific form of the one-sided score.

\begin{corollary}[MDR and SDR control for interval targets]
\label{cor:two_sided_risk}
Consider either two-sided calibration scheme described above, and let $\widehat S_{\mathcal E,\delta}$ denote the selection set obtained with adjustment $\delta\ge0$. We retain the conditioning and target assumptions of the corresponding one-sided setting, with thresholds replaced by intervals. Assume that the fitted scores are real-valued and the interval losses take values in $[0,1]$. If the calibration and test observations are exchangeable, then, for every $n_2\ge1$ and $m\ge1$,
$$
\mathrm{MDR} \bigl(\widehat S_{\mathrm{MDR},\delta}\bigr) \le q,
\qquad \delta\ge1-q.
$$
The unadjusted procedure satisfies
$$
\mathrm{MDR} \bigl(\widehat S_{\mathrm{MDR},0}\bigr) \le q+\frac{1-q}{n_2+1}.
$$
Furthermore, if the observations are i.i.d. and Assumption~\ref{assump:unec_reg} holds for the joint distribution of the interval score and loss, separately for each target in the external setting, then, for any fixed $\delta\ge0$ and $m$,
$$
\limsup_{n_2\to\infty}\mathrm{SDR}\bigl(\widehat S_{\mathrm{SDR},\delta}\bigr) \le q.
$$
\end{corollary}

\begin{proof}
The proof of Theorem~\ref{thm:finite_mdr} applies to any fixed real-valued score and bounded loss. It therefore applies directly to the common or paired interval procedure. For external intervals, apply the same argument separately to each fixed target $(a_j,b_j)$ and average the resulting expected-loss bounds over $j$. The unadjusted MDR bound follows in the same way from Corollary~\ref{cor:asymptotic_mdr}.

For $\delta=0$, the SDR conclusions follow from the proofs of Theorems~\ref{thm:unec_asymptotic} and~\ref{thm:external_sdr}, respectively, after substituting the interval scores and losses. For any fixed $\delta>0$, the adjustment to the empirical SDR is
$$
\frac{\delta}{1\vee\sum_{i\in I_{\mathrm{cal}}}\mathbbm{1}\{Z_i\ge z\}},
$$
with $Z_i$ replaced by $Z_{i,j}$ in the external setting. This term converges uniformly to zero on regions where the population selection probability is bounded away from zero. The same feasibility and convergence arguments therefore yield the stated asymptotic SDR bound.
\end{proof}

If the conditional assumptions hold almost surely, the corresponding unconditional risk bounds follow by averaging and boundedness.

\section{Technical Proofs}\label{app:proofs}

\subsection{Proof of theorem~\ref{thm:unec_asymptotic}}
\label{app:asym_sdr_proof}
\paragraph{Regularity conditions}
We first state the regularity conditions used in the asymptotic analysis. Throughout, we work under the conditional law given the training data and fitting randomness, so that the score function is fixed. The quantities $F$, $G$, and $Q$ are defined in Section~\ref{subsec:asymptotic}.
\begin{assump}[Regularity conditions]
\label{assump:unec_reg}
Under the conditional law associated with the fixed training stage:
\begin{enumerate}
    \item The cumulative distribution function of $Z$ is continuous.
    \item $\mathbb{E}[L]>q$, and the set $\{z\in\mathbb{R}:F(z)>0,\ Q(z)\le q\}$ is nonempty.
    \item Define $z^*=\inf\{z\in\mathbb{R}:F(z)>0,\ Q(z)\le q\}.$ For any $\epsilon>0$, there exists $z_\epsilon\in(z^*,z^*+\epsilon)$ such that $F(z_\epsilon)>0$ and $Q(z_\epsilon)<q$.
\end{enumerate}
\end{assump}

The first condition, together with the boundedness of the loss, ensures the continuity of $F(z)$ and $G(z)$, and hence the continuity of $Q(z)=G(z)/F(z)$ whenever $F(z)>0$. The second condition restricts attention to the nontrivial regime in which selecting all observations exceeds the target risk, while at least one population cutoff satisfies the risk constraint. The last condition requires strictly feasible cutoffs arbitrarily close to $z^*$ from above. It excludes a first feasible cutoff that merely touches the target level without nearby strict feasibility, allowing empirical calibration to approximate the population cutoff despite sampling fluctuations. This is a local condition and does not require $Q(z)$ to be globally monotone.

\begin{proof}
Fix a realization of the training stage for which Assumption~\ref{assump:unec_reg} holds. All probabilities and expectations in this proof are taken under the resulting conditional law, following the convention in Section~\ref{subsec:asymptotic}. Thus, conditioning on $\mathcal{D}_{\mathrm{cal}}$ or $\widehat{Z}_{n_2}$ below is additional to the training information already held fixed.

Let $\widehat{F}_{n_2}(z) = \frac{1}{n_2} \sum_{i \in I_{\mathrm{cal}}} \mathbbm{1}\{Z_i \ge z\}$ and $\widehat{G}_{n_2}(z) = \frac{1}{n_2} \sum_{i \in I_{\mathrm{cal}}} L_i \mathbbm{1}\{Z_i \ge z\}$. 
Because the class of indicator functions over half-spaces $\mathcal{F} = \{ \mathbbm{1}\{Z \ge z\} : z \in \mathbb{R} \}$ has a Vapnik-Chervonenkis (VC) dimension of 1, it forms a uniform Glivenko-Cantelli class. Moreover, because $0\le L\le 1$, the weighted class $\mathcal G=\{(Z,L)\mapsto L\mathbbm{1}\{Z\ge z\}:z\in\mathbb R\}$ is a bounded VC-subgraph class and is also Glivenko-Cantelli. By the Glivenko-Cantelli theorem, the empirical proportions converge almost surely to their population counterparts uniformly over $z \in \mathbb{R}$:
\begin{equation}\label{eq:gc_convergence}
    \sup_{z \in \mathbb{R}} |\widehat{F}_{n_2}(z) - F(z)| \xrightarrow{a.s.} 0, \quad \text{and} \quad \sup_{z \in \mathbb{R}} |\widehat{G}_{n_2}(z) - G(z)| \xrightarrow{a.s.} 0.
\end{equation}

By Assumption~\ref{assump:unec_reg}, there exists $z_0$ such that $F(z_0) > 0$ and $Q(z_0) < q$. Pick $\delta > 0$ such that $F(z_0) \ge \delta$. Let $\Omega_\delta = \{ z \in \mathbb{R} : F(z) \ge \delta \}$. On $\Omega_\delta$, the denominator $F(z)$ is strictly bounded away from zero. Therefore, the uniform convergence of the numerator and denominator, combined with the standard algebra of uniformly convergent limits (since the denominator is strictly bounded away from zero by $\delta$):
\begin{equation}
    \sup_{z \in \Omega_\delta} |\widehat{\textnormal{SDR}}_{n_2}(z) - Q(z)| \xrightarrow{a.s.} 0.
    \label{eq:Qn_Q}
\end{equation}

We first relate a strictly feasible population cutoff to the empirical candidate set. For any fixed $t$ satisfying $F(t)>0$ and $Q(t)<q$, define $k_{n_2}(t) = \sum_{i\in I_{\mathrm{cal}}}\mathbbm{1}\{Z_i\ge t\}, s_{n_2}(t) = \max\{Z_i:i\in I_{\mathrm{cal}},\ Z_i\le t\},$ whenever the set defining $s_{n_2}(t)$ is nonempty. Since $Q(t)<q<\mathbb{E}[L]$, we must have $\mathbb{P}(Z<t)>0$; otherwise $F(t)=1$ and $Q(t)=\mathbb{E}[L]$. Thus $s_{n_2}(t)$ exists eventually almost surely. Moreover, $$\frac{k_{n_2}(t)}{n_2}\xrightarrow{\mathrm{a.s.}}F(t)>0.$$

By continuity of the score distribution, the calibration scores are distinct almost surely. Moving the cutoff from $t$ to $s_{n_2}(t)$ therefore adds at most one calibration observation. Since the losses lie in $[0,1]$, $$\left|\widehat{\mathrm{SDR}}_{n_2}(s_{n_2}(t)) - \widehat{\mathrm{SDR}}_{n_2}(t)\right| \le \frac{1}{k_{n_2}(t)+1}\xrightarrow{\mathrm{a.s.}}0.$$ By Eq.~\eqref{eq:gc_convergence} and $F(t)>0$, $$\widehat{\mathrm{SDR}}_{n_2}(t)\xrightarrow{\mathrm{a.s.}}Q(t)<q.$$ Consequently, $s_{n_2}(t)$ is a feasible empirical candidate for all sufficiently large $n_2$, almost surely.

Applying this construction with $t=z_0$ gives $$\widehat{Z}_{n_2} \le s_{n_2}(z_0) \le z_0$$ eventually almost surely. Therefore, $$F(\widehat{Z}_{n_2})\ge F(z_0)\ge\delta,$$ so $\widehat{Z}_{n_2}\in\Omega_\delta$ eventually almost surely. Combining Eq.~\eqref{eq:Qn_Q} with empirical feasibility yields the following bound:
\begin{equation}\label{eq:pop_risk_bound}
    Q(\widehat{Z}_{n_2}) \le \widehat{\textnormal{SDR}}_{n_2}(\widehat{Z}_{n_2}) + o_p(1) \le q + o_p(1).
\end{equation}

Conditional on $\mathcal{D}_{\mathrm{cal}}$, according to the definition of conditional probability, for $j \in \mathcal{D}_{\mathrm{test}}$,
\begin{equation}
\begin{aligned}
    \mathbb{E}[L_{n+j} \cdot \mathbbm{1}\{Z_j^{\mathrm{test}}\ge \widehat{Z}_{n_2}\} \mid \mathcal{D}_{\mathrm{cal}}] = &\mathbb{E}[L_{n+j} \mid Z_j^{\mathrm{test}} \ge \widehat{Z}_{n_2}, \mathcal{D}_{\mathrm{cal}}] \\
    &\times \mathbb{E}[\mathbbm{1}\{Z_j^{\mathrm{test}} \ge \widehat{Z}_{n_2}\} \mid \mathcal{D}_{\mathrm{cal}}]
\end{aligned}
\end{equation}
Denote $\widehat{S}=\{j \in I_{\mathrm{test}}:Z_j^{\mathrm{test}} \ge \widehat{Z}_{n_2}\}$ as the output of UNEC summarized in Algorithm~\ref{algo:UNEC_constant}. As a result, the $\textnormal{SDR}_{\textnormal{test}}(\widehat{Z}_{n_2})$ can be written as
\begin{equation}
\begin{aligned}
    \textnormal{SDR}_{\textnormal{test}}(\widehat{Z}_{n_2}) &= \mathbb{E}\left[\mathbb{E}\left[\frac{\sum_{j=1}^{m}{L_{n+j} \cdot \mathbbm{1}\{Z_j^{\mathrm{test}} \ge \widehat{Z}_{n_2}\}}}{\max(1,\sum_{j=1}^{m}\mathbbm{1}\{Z_j^{\mathrm{test}} \ge \widehat{Z}_{n_2}\})} \middle| \widehat{S},\mathcal{D}_{\mathrm{cal}}\right]\right]\\
    &= \mathbb{E}\left[\mathbb{E}\left[\frac{\sum_{j=1}^{m}{L_{n+j} \cdot \mathbbm{1}\{j \in \widehat{S}\}}}{\max(1,|\widehat{S}|)} \middle| \widehat{S},\mathcal{D}_{\mathrm{cal}}\right]\right]\\
    &= \mathbb{E}\left[\frac{|\widehat{S}| \times Q(\widehat{Z}_{n_2})}{\max(1,|\widehat{S}|)}\right]\\
    &\le \mathbb{E}\left[Q(\widehat{Z}_{n_2})\right].
\end{aligned}
\end{equation}
Let the excess error be $E = \max(Q(\widehat{Z}_{n_2}) - q, 0)$. Because $Q(\widehat{Z}_{n_2}) \le q + o_p(1)$, we have $E \xrightarrow{p} 0$. Since $Q(z)$ is bounded in $[0,1]$, $E$ is uniformly integrable. By the Bounded Convergence Theorem, $\mathbb{E}[E] \to 0$. Taking expectations yields $\mathbb{E}[Q(\widehat{Z}_{n_2})] \le q + \mathbb{E}[E]$, and taking the superior limit directly yields:
\begin{equation}
    \limsup_{n_2 \to \infty} \textnormal{SDR}_{\textnormal{test}}(\widehat{Z}_{n_2}) \le  \limsup_{n_2 \to \infty} \mathbb{E}[Q(\widehat{Z}_{n_2})] \le q.
\end{equation}

We next derive an explicit expression for the conditional test-set SDR. Since the fitted score function is fixed and the calibration sample is independent of the test sample under the conditional law, the cutoff $\widehat{Z}_{n_2}$ is independent of the test observations under this law. Consequently,
$$
|\widehat{S}|\mid\widehat{Z}_{n_2} \sim \operatorname{Binom} \left(m,F(\widehat{Z}_{n_2})\right),
$$
and hence
$$
\mathbb{P}\left(|\widehat{S}|>0\mid\widehat{Z}_{n_2}\right) = 1-\left\{1-F(\widehat{Z}_{n_2})\right\}^{m}.
$$
Using the conditional mean loss among selected observations, we obtain
$$
\begin{aligned}
\mathrm{SDR}_{\mathrm{test}}(\widehat{Z}_{n_2})
&= \mathbb{E}\left[Q(\widehat{Z}_{n_2})\mathbbm{1}\{|\widehat{S}|>0\}\right]\\
&=\mathbb{E}\left[Q(\widehat{Z}_{n_2})\mathbb{P}\left(|\widehat{S}|>0\mid\widehat{Z}_{n_2}\right)\right]\\
&=\mathbb{E}\left[Q(\widehat{Z}_{n_2})\left\{1-\left(1-F(\widehat{Z}_{n_2})\right)^m\right\}\right].
\end{aligned}
$$
For completeness, we set $Q(z)=0$ whenever $F(z)=0$, in which case no test observation is selected with positive
probability.

We next establish consistency of the calibrated cutoff. Since $Z$ is real-valued and $0\le L\le1$,
$$F(z)\to1, \qquad G(z)\to\mathbb{E}[L],\qquad Q(z)\to\mathbb{E}[L]>q \quad\text{as }z\to-\infty.$$
The population feasible set is therefore bounded below. Its nonemptiness implies that $z^*\in\mathbb{R}$. Since $z_0$ is strictly feasible, $z^*\le z_0$ and $F(z^*)\ge F(z_0)\ge\delta>0.$

Fix $\epsilon>0$, and for every $z\le z^*-\epsilon$, we have $F(z)>0$ and $Q(z)>q$ by the definition of $z^*$. The limit $Q(z)\to\mathbb{E}[L]>q$ separates $Q(z)$ from $q$ on a sufficiently far-left tail. Together with continuity of $Q(z)$ and the strict inequality $Q(z)>q$ on the remaining compact interval, this yields $\eta_\epsilon :=\inf_{z\le z^*-\epsilon}\{Q(z)-q\}>0.$ For all such $z$, $F(z)\ge F(z_0)\ge\delta,$ so this region is contained in $\Omega_\delta$. By Eq.~\eqref{eq:Qn_Q}, eventually $$\widehat{\mathrm{SDR}}_{n_2}(z) > q+\frac{\eta_\epsilon}{2} \qquad \text{for all }z\le z^*-\epsilon,$$ almost surely. Therefore, for sufficiently large $n_2$, $\widehat{Z}_{n_2}>z^*-\epsilon.$

For the upper bound, choose $z_\epsilon\in(z^*,z^*+\epsilon)$ as in Assumption~\ref{assump:unec_reg}. The preceding candidate construction, now applied with $t=z_\epsilon$, shows that $s_{n_2}(z_\epsilon)$ is eventually a feasible empirical candidate. Consequently, for sufficiently large $n_2$,
$$
\widehat{Z}_{n_2} \le s_{n_2}(z_\epsilon) \le z_\epsilon < z^*+\epsilon.
$$
Since $\epsilon > 0$ is arbitrary, it follows that, as $n_2 \to \infty,$ $\widehat{Z}_{n_2}\xrightarrow{\mathrm{a.s.}}z^*.$

We also verify that $Q(z^*)=q$. Continuity of $Q$ near $z^*$ and the existence of strictly feasible points approaching $z^*$ from the right imply $Q(z^*)\le q$. If $Q(z^*)<q$, continuity would imply the existence of a feasible cutoff strictly below $z^*$, contradicting its definition. Thus $Q(z^*)=q$.

Since $F$ and $G$ are continuous and $F(z^*)>0$, cutoff consistency gives 
$$
F(\widehat{Z}_{n_2})\xrightarrow{\mathrm{a.s.}}\beta=F(z^*), \qquad Q(\widehat{Z}_{n_2})\xrightarrow{\mathrm{a.s.}}q.
$$
Using the exact conditional SDR identity established above, and noting that its integrand lies in $[0,1]$, the bounded convergence theorem yields
$$\lim_{n_2\to\infty} \mathrm{SDR}_{\mathrm{test}}(\widehat{Z}_{n_2}) = q\{1-(1-\beta)^m\}.$$

For the power limit, assume $\gamma>0$. The continuity of the score distribution implies continuity of $z\mapsto\mathbb{P}(Z\ge z,Y\in\mathcal{R}_1)$, and hence of $Pow(z)$. Therefore, $Pow(\widehat{Z}_{n_2})\xrightarrow{\mathrm{a.s.}}Pow(z^*)$. Similarly, we can also show the expression of power by decomposing it. Let $K = \sum_{j=1}^m \mathbbm{1}\{Y_{j} \in \mathcal{R}_1\},$ we get
$$
\begin{aligned}
    &\lim_{n_2\to\infty} \textnormal{Power}_{\textnormal{test}}(\widehat{Z}_{n_2}) \\
    =& \lim_{n_2\to\infty}\mathbb{E}\left[\frac{\sum_{j=1}^m \mathbbm{1}\{Z_j \ge \widehat{Z}_{n_2}, Y_{j} \in \mathcal{R}_1\}}{\max\left(1, \sum_{j=1}^m \mathbbm{1}\{Y_{j} \in \mathcal{R}_1\} \right)}\right]\\
    =& \lim_{n_2\to\infty} \mathbb{E} \left[\mathbb{E} \left[\frac{\sum_{j=1}^{m}{\mathbbm{1}\{Z_j \ge \widehat{Z}_{n_2}, Y_{j} \in \mathcal{R}_1\}}}{\max(1,\sum_{j=1}^{m}{\mathbbm{1}\{Y_{j} \in \mathcal{R}_1}\})} \middle| \mathbbm{1}\{Y_{j} \in \mathcal{R}_1\},j=1\cdots,m,\mathcal{D}_{\mathrm{cal}}\right]\right]\\
    =& \lim_{n_2\to\infty} \mathbb{E}\left[\frac{K \times Pow(\widehat{Z}_{n_2})}{\max(1,K)}\right]\\
    =& \lim_{n_2\to\infty}\mathbb{E}\left[Pow(\widehat{Z}_{n_2}) \times \mathbbm{1}(K>0)\right]\\
    =& \lim_{n_2\to\infty}\mathbb{E}\left[Pow(\widehat{Z}_{n_2})\right] \times \mathbb{P}(K>0)\\
    =& \lim_{n_2\to\infty}\mathbb{E}\left[Pow(\widehat{Z}_{n_2})\right]\left\{1-\bigl(1-\gamma\bigr)^m\right\}\\
    =& Pow(z^*)\left\{1-\bigl(1-\gamma\bigr)^m\right\}.
\end{aligned}
$$

This completes the proof under the conditional law associated with the fixed training data and fitting randomness.
\end{proof}

\subsection{Proof of theorem~\ref{thm:finite_mdr}}
\begin{proof}
We first condition on the training data and fitting randomness, so that the fitted score function is fixed. Until the final averaging step, all expectations are taken under this conditional law.

Let $R_i(z)=L_i\mathbbm{1}\{Z_i\ge z\}, i\in I_{\mathrm{cal}},$ and $R_{n+j}(z) = L_{n+j,j}\mathbbm{1}\{Z(X_{n+j};c)\ge z\}, j\in\{1,\ldots,m\}.$ Write
$$
\widehat{R}_{n_2}(z) = \frac{1}{n_2}\sum_{i\in I_{\mathrm{cal}}}R_i(z).
$$
These loss functions take values in $[0,1]$ and are non-increasing in $z$. We set $R_i(+\infty)=0$ for every observation considered below.

Following the exchangeability proof strategy of \citet{angelopoulos2024conformal}, consider the auxiliary cutoff
$$
\dot Z = \min\left\{z\in\mathcal Z_{\mathrm{cand}}\cup\{Z(X_{n+j};c)\}: \frac{n_2\widehat{R}_{n_2}(z)+R_{n+j}(z)}{n_2+1} \le q \right\},
$$
if the feasible set is nonempty, and set $\dot Z=+\infty$ otherwise.

Both the candidate set and the total loss in this definition are invariant under permutations of the $n_2+1$ observations. Thus, $\dot Z$ is a symmetric function of these observations. By conditional exchangeability and the definition of $\dot Z$,
$$
\mathbb E\!\left[R_{n+j}(\dot Z)\right] = \frac{1}{n_2+1}\mathbb E\!\left[\sum_{i\in I_{\mathrm{cal}}}R_i(\dot Z)+R_{n+j}(\dot Z)\right]\le q.
$$
This inequality also holds when $\dot Z=+\infty$, since all loss contributions are then zero.

If $\widehat{Z}_{\mathrm{MDR},\delta}<+\infty$, empirical feasibility gives
$$
n_2\widehat{R}_{n_2} \bigl(\widehat{Z}_{\mathrm{MDR},\delta}\bigr)+\delta \le n_2q.
$$
Since $R_{n+j}(z)\le1$ and $\delta\ge1-q$,
$$
\frac{n_2\widehat{R}_{n_2}\bigl(\widehat{Z}_{\mathrm{MDR},\delta}\bigr) + R_{n+j} \bigl(\widehat{Z}_{\mathrm{MDR},\delta}\bigr)}{n_2+1}\le\frac{n_2q-\delta+1}{n_2+1}\le q.
$$
Moreover, $\widehat{Z}_{\mathrm{MDR},\delta}\in \mathcal Z_{\mathrm{cand}}$, so it is a feasible candidate in the definition of $\dot Z$. Therefore, $\widehat{Z}_{\mathrm{MDR},\delta}\ge\dot Z$. The same ordering holds when $\widehat{Z}_{\mathrm{MDR},\delta}=+\infty$.

Since $R_{n+j}(z)$ is non-increasing in $z$,
$$
R_{n+j} \bigl(\widehat{Z}_{\mathrm{MDR},\delta}\bigr) \le R_{n+j}(\dot Z).
$$
Taking expectations gives
$$
\mathbb E\!\left[R_{n+j}\bigl(\widehat{Z}_{\mathrm{MDR},\delta}\bigr)\right]\le q.
$$
Since $j$ was arbitrary, this conditional bound holds for every test observation. Averaging over the training data and fitting randomness gives the corresponding unconditional bound. Finally, averaging over $j=1,\ldots,m$ yields $\mathrm{MDR}(\widehat{S}_{\mathrm{MDR},\delta})\le q$.
\end{proof}

\subsection{Asymptotic MDR control for the unadjusted procedure}
\label{app:asymptotic_mdr}

\begin{corollary}[Asymptotic MDR control]
\label{cor:asymptotic_mdr}
Under the common-threshold setting and the conditional exchangeability and bounded-loss assumptions of Theorem~\ref{thm:finite_mdr}, let $\widehat{S}_{\mathrm{MDR}}$ be the selection set obtained by the unadjusted MDR procedure at level $q\in(0,1)$. For every $n_2\ge1$ and $m\ge1$,
$$
\mathrm{MDR}(\widehat{S}_{\mathrm{MDR}}) \le \frac{n_2q+1}{n_2+1} = q+\frac{1-q}{n_2+1}.
$$
Consequently,
$$
\limsup_{n_2\to\infty} \mathrm{MDR}(\widehat{S}_{\mathrm{MDR}}) \le q.
$$
These bounds hold both conditional on the training stage and unconditionally.
\end{corollary}

\begin{proof}
Set
$$
q_{n_2}=\frac{n_2q+1}{n_2+1}\in(0,1).
$$
Since $q_{n_2}-\frac{1-q_{n_2}}{n_2}=q,$ the unadjusted empirical constraint $\widehat{\mathrm{MDR}}_{n_2}(z)\le q$ is equivalent to
$$
\widehat{\mathrm{MDR}}_{n_2}(z)+\frac{1-q_{n_2}}{n_2} \le q_{n_2}.
$$
Thus, the unadjusted MDR procedure at level $q$ coincides with the adjusted MDR procedure at level $q_{n_2}$ with adjustment $\delta=1-q_{n_2}$. Their feasible candidate sets, selected cutoffs, and empty-set conventions are identical.

Applying Theorem~\ref{thm:finite_mdr} at level $q_{n_2}$ gives
$$
\mathrm{MDR}(\widehat{S}_{\mathrm{MDR}}) \le q_{n_2}.
$$
The asymptotic conclusion follows from $q_{n_2}\to q$.
\end{proof}

\subsection{Proofs for asymptotic power optimality}
\label{app:power_optimality}
We first formalize the oracle optimization problem and then connect the data-driven UNEC procedure to this oracle benchmark. We consider the one-sided binary-loss setting with $\mathcal R_0=(-\infty,c)$ and $\mathcal R_1=[c,\infty)$. Let
$$
Z^o(X)=\frac{\mu(X)-c}{\sigma(X)}
$$
and, for $z\in\mathbb R$, define
$$
F^o(z)=\mathbb P\{Z^o(X)\ge z\}, \qquad G^o(z)=\mathbb P\{Z^o(X)\ge z,\ Y<c\},
$$
$$
Q^o(z)=\frac{G^o(z)}{F^o(z)} \quad\text{for }F^o(z)>0, \qquad Pow^o(z)=\mathbb P\{Z^o(X)\ge z\mid Y\ge c\}.
$$

\begin{assump}[Regularity conditions for power optimality]
\label{assump:opt_power}
We assume:
\begin{enumerate}
    \item The cumulative distribution function of $Z^o(X)$ is continuous.
    \item $q<\mathbb P(Y<c)<1$, and the set $\{z\in\mathbb R:F^o(z)>0,\ Q^o(z)\le q\}$ is nonempty.
    \item Define $z^o=\inf\{z\in\mathbb R:F^o(z)>0,\ Q^o(z)\le q\}.$ For any $\epsilon>0$, there exists $z_\epsilon^o\in(z^o,z^o+\epsilon)$ such that $F^o(z_\epsilon^o)>0,$ and $Q^o(z_\epsilon^o)<q.$
\end{enumerate}
\end{assump}

The first condition ensures regularity of oracle threshold sets, while the second and third conditions guarantee a nontrivial and locally well-identified oracle cutoff.

The first step is to show that the oracle standardized margin induces the best possible ordering of observations. For a measurable observation-wise rule $S$, write
$$
Pow(S)=\mathbb P\{S(X)=1\mid Y\ge c\}, \qquad Q(S)=\mathbb P\{Y<c\mid S(X)=1\}.
$$
\begin{colorboxyellow}
\begin{lemma}[Oracle threshold dominance]
\label{lem:oracle_ranking}
Under Assumption~\ref{assump:opt_power} and the location--scale model, for any measurable observation-wise selection rule $S(X)\in\{0,1\}$ with $p_S=\mathbb P\{S(X)=1\}\in(0,1)$, there exists an oracle threshold rule $S^o_{z_S}(X)=\mathbbm{1}\{Z^o(X)\ge z_S\},$ with $\mathbb P\{S^o_{z_S}(X)=1\}=p_S,$ such that $Pow(S^o_{z_S})\ge Pow(S)$ and $Q(S^o_{z_S})\le Q(S).$ Consequently, for every test size $m$,
$$
\mathrm{Power}_m(S^o_{z_S}) \ge \mathrm{Power}_m(S),
\qquad
\mathrm{SDR}_m(S^o_{z_S}) \le \mathrm{SDR}_m(S).
$$
\end{lemma}
\end{colorboxyellow}

\begin{proof}
Let $\eta(x)=\mathbb P(Y\ge c\mid X=x).$ Under the location--scale model,
$$
\eta(x) = \mathbb P\left\{\epsilon\ge \frac{c-\mu(x)}{\sigma(x)}\right\} = 1-F_\epsilon\{-Z^o(x)\},
$$
so $\eta(x)$ is non-decreasing in $Z^o(x)$.

Let $A=\{x:S(x)=1\}, B=\{x:Z^o(x)\ge z_S\},$ where $z_S$ is chosen so that
$$
\mathbb P(X\in B)=\mathbb P(X\in A)=p_S.
$$
Such a threshold exists because the distribution of $Z^o(X)$ is continuous.

Since $B$ is an upper level set of $Z^o$, for $x\in B\setminus A$ and $x'\in A\setminus B$,
$$
Z^o(x)\ge z_S>Z^o(x'),
$$
and hence $\eta(x)\ge\eta(x').$ Moreover, $\mathbb P(X\in B\setminus A)=\mathbb P(X\in A\setminus B),$ because $A$ and $B$ have the same probability. It follows that
$$
\mathbb E\{\eta(X)\mathbbm{1}(X\in B)\} \ge \mathbb E\{\eta(X)\mathbbm{1}(X\in A)\}.
$$
Since $\mathbb P\{S(X)=1,Y\ge c\}=\mathbb E\{\eta(X)S(X)\},$ dividing by $\gamma=\mathbb P(Y\ge c)$ gives
$$
Pow(S^o_{z_S})\ge Pow(S).
$$

Because both rules have selection probability $p_S$, $\mathbb P\{S(X)=1,Y<c\}=p_S-\mathbb E\{\eta(X)S(X)\}.$ Therefore,
$$
\mathbb P\{S^o_{z_S}(X)=1,Y<c\} \le \mathbb P\{S(X)=1,Y<c\},
$$
and division by the common selection probability $p_S$ yields $Q(S^o_{z_S})\le Q(S).$

It remains to translate these population inequalities to a test sample of size $m$. For any observation-wise rule $S$, let $R=\sum_{j=1}^m S(X_{n+j}).$ Conditional on the selection indicators, whenever $R>0$, the expected average false-selection loss among the selected observations equals $Q(S)$. Hence
$$
\mathrm{SDR}_m(S) = Q(S)\mathbb P(R>0) = Q(S)\{1-(1-p_S)^m\}.
$$
Since $S$ and $S^o_{z_S}$ have the same selection probability $p_S$, the preceding inequality for $Q$ implies $\mathrm{SDR}_m(S^o_{z_S}) \le \mathrm{SDR}_m(S).$

Similarly, if $M=\sum_{j=1}^m\mathbbm{1}\{Y_{n+j}\ge c\},$ then
$$
\mathrm{Power}_m(S) = Pow(S)\mathbb P(M>0) = Pow(S)\{1-(1-\gamma)^m\}.
$$
The multiplicative factor does not depend on the selection rule, so $\mathrm{Power}_m(S^o_{z_S}) \ge \mathrm{Power}_m(S).$
\end{proof}

Lemma~\ref{lem:oracle_ranking} shows that, for any observation-wise selection rule, there exists an oracle threshold rule based on $Z^o$ with the same selection probability, no smaller power, and no larger conditional false-selection risk. Hence, when searching for the finite-sample power-optimal rule under an SDR constraint, it is sufficient to restrict attention to threshold rules based on $Z^o$. We next characterize the optimal threshold within this oracle family and show that its finite-$m$ cutoff converges to the population cutoff $z^o$ as the test sample size grows.

\begin{colorboxyellow}
\begin{lemma}[Oracle cutoff characterization and convergence]
\label{lem:oracle_cutoff}
Under Assumption~\ref{assump:opt_power}, for a test sample of size $m$, define $R_m^o(z)=Q^o(z)\left[1-\{1-F^o(z)\}^m\right],$ and $z_m^o=\inf\left\{z:F^o(z)>0,\ R_m^o(z)\le q\right\}.$ Then the oracle threshold rule $S_{z_m^o}^o(X)=\mathbbm{1}\{Z^o(X)\ge z_m^o\}$ maximizes test power among all measurable observation-wise selection rules satisfying $\mathrm{SDR}_m\le q$. Moreover,
$$
z_m^o\longrightarrow z^o, \qquad \text{as }m\to\infty.
$$
\end{lemma}
\end{colorboxyellow}

\begin{proof}
For an oracle threshold rule $S_z^o(X)=\mathbbm{1}\{Z^o(X)\ge z\}$, the population selection probability is $F^o(z)$ and the conditional false-selection risk is $Q^o(z)$. By the finite-test representation in Lemma~\ref{lem:oracle_ranking},
$$
\mathrm{SDR}_m(S_z^o)=Q^o(z)\left[1-\{1-F^o(z)\}^m\right]=R_m^o(z).
$$
Similarly, writing $\gamma=\mathbb P(Y\ge c),$ the corresponding test power is
$$
\mathrm{Power}_m(S_z^o)=Pow^o(z)\{1-(1-\gamma)^m\}.
$$

Lemma~\ref{lem:oracle_ranking} shows that when maximizing test power subject to $\mathrm{SDR}_m\le q$, it is sufficient to consider the family $\{S_z^o:z\in\mathbb R\}$. The boundary cases do not affect the optimum. A rule with $p_S=1$ is infeasible because $\mathbb P(Y<c)>q$, whereas a rule with $p_S=0$ has zero power and is dominated by a nonempty feasible oracle threshold rule guaranteed by Assumption~\ref{assump:opt_power}.

Since $\eta(X)$ is non-decreasing in $Z^o(X)$, $Q^o(z)=1-\mathbb E\{\eta(X)\mid Z^o(X)\ge z\}$ is non-increasing in $z$. Moreover, $F^o(z)$ and $Pow^o(z)$ are non-increasing in $z$. Therefore, $R_m^o(z)$ is non-increasing in $z$. Thus the smallest threshold satisfying $R_m^o(z)\le q$ yields the largest power among all feasible oracle threshold rules. By continuity and Assumption~\ref{assump:opt_power}, the infimum defining $z_m^o$ is feasible, so $S_{z_m^o}^o$ is power-optimal under the finite-$m$ SDR constraint.

It remains to show that $z_m^o\to z^o$. By the definition of $z^o$ and Assumption~\ref{assump:opt_power}, for any $\epsilon>0,$
$$
F^o(z^o-\epsilon)>0,
\qquad
Q^o(z^o-\epsilon)>q.
$$
Since $1-\{1-F^o(z^o-\epsilon)\}^m \longrightarrow 1,$ we have
$$
R_m^o(z^o-\epsilon) \longrightarrow Q^o(z^o-\epsilon)>q.
$$
Hence, for all sufficiently large $m$, $R_m^o(z^o-\epsilon)>q.$ Since $R_m^o$ is non-increasing, no threshold $z\le z^o-\epsilon$ is feasible, and therefore $z_m^o>z^o-\epsilon.$

On the other hand, Assumption~\ref{assump:opt_power} guarantees the existence of some $z_\epsilon^o\in(z^o,z^o+\epsilon)$ such that $F^o(z_\epsilon^o)>0,$ and $Q^o(z_\epsilon^o)<q.$ Because
$$
R_m^o(z_\epsilon^o) = Q^o(z_\epsilon^o)\left[1-\{1-F^o(z_\epsilon^o)\}^m\right] \le Q^o(z_\epsilon^o) < q,
$$
$z_\epsilon^o$ is feasible for every $m$. Hence, $z_m^o \le z_\epsilon^o < z^o+\epsilon.$

Combining the two bounds, for all sufficiently large $m$,
$$
z^o-\epsilon < z_m^o < z^o+\epsilon.
$$
Since $\epsilon>0$ is arbitrary, $z_m^o\longrightarrow z^o.$
\end{proof}

Lemmas~\ref{lem:oracle_ranking} and~\ref{lem:oracle_cutoff} together characterize the finite-sample oracle problem: it is sufficient to consider threshold rules based on $Z^o$, and the corresponding SDR-optimal cutoff converges to the population cutoff $z^o$ as the test sample size grows. We finally show that the data-driven UNEC procedure converges to this oracle limit in Theorem~\ref{thm:opt_power}.

\noindent\textbf{Restatement of Theorem~\ref{thm:opt_power}.}
Suppose the location--scale model above holds and the regularity conditions in Assumption~\ref{assump:opt_power} are satisfied. Further assume $\sup_x|\hat\mu_{n_0}(x)-\mu(x)|\overset{p}\longrightarrow0, \sup_x|\hat\sigma_{n_1}(x)-\sigma(x)|\overset{p}\longrightarrow0, \sup_x|\mu(x)-c|<\infty,$ and $\inf_x\sigma(x)\ge \underline{\sigma}>0$. Then, UNEC asymptotically attains the oracle power envelope under SDR/FDR control among observation-wise selection rules.

\begin{proof}
We first show that the fitted UNEC score converges uniformly to its oracle counterpart. By definition,
$$
Z(x)-Z^o(x) = \frac{\hat\mu_{n_0}(x)-\mu(x)}{\hat\sigma_{n_1}(x)} + \{\mu(x)-c\}\frac{\sigma(x)-\hat\sigma_{n_1}(x)}{\hat\sigma_{n_1}(x)\sigma(x)}.
$$
Since $\inf_x\sigma(x)\ge\underline{\sigma}>0$ and $\hat\sigma_{n_1}$ converges uniformly to $\sigma$, with probability tending to one, $\inf_x\hat\sigma_{n_1}(x)\ge\underline{\sigma}/2.$ Under Assumption~\ref{assump:opt_power}, together with the uniform consistency of $\hat\mu_{n_0}$ and $\hat\sigma_{n_1}$ and the boundedness of $\mu(x)-c$, this gives
$$
\Delta_{n_0,n_1} := \sup_x|Z(x)-Z^o(x)|\overset{p}\longrightarrow0.
$$

We next show uniform convergence of the population risk and power. For any $r>0$, on the event $\Delta_{n_0,n_1}\le r$, we have $\{Z^o\ge z+r\} \subseteq \{Z\ge z\} \subseteq \{Z^o\ge z-r\}$ for every $z \in \mathbb{R}$. Hence
$$
F^o(z+r)\le F(z)\le F^o(z-r).
$$
Because the distribution function of $Z^o$ is continuous, it is uniformly continuous, and therefore 
$$
\sup_z|F(z)-F^o(z)| \overset{p}\longrightarrow0.
$$
Similarly, we can obtain 
$$
\sup_z|G(z)-G^o(z)| \overset{p}\longrightarrow0, \qquad \sup_z|Pow(z)-Pow^o(z)|\overset{p}\longrightarrow0.
$$

We next show that $z^*\overset{p}\longrightarrow z^o.$ For any $\epsilon>0$, by the definition of $z^o$ and Assumption~\ref{assump:opt_power}, $F^o(z^o-\epsilon)>0,$ and $Q^o(z^o-\epsilon)>q.$ Moreover, there exists $z_\epsilon^o\in(z^o,z^o+\epsilon)$ such that $F^o(z_\epsilon^o)>0,$ and $Q^o(z_\epsilon^o)<q.$ Since $F^o$ is bounded away from zero on $(-\infty,z_\epsilon^o]$, the uniform convergence of $F$ and $G$ implies uniform convergence of $Q=G/F$ to $Q^o=G^o/F^o$ over this region. Consequently, with probability tending to one,
$$
Q(z)>q, \qquad \text{for all }z\le z^o-\epsilon,
$$
while $Q(z_\epsilon^o)<q.$ It follows from the definition of $z^*$ that, with probability tending to one, $z^o-\epsilon<z^*\le z_\epsilon^o<z^o+\epsilon.$ Therefore, $z^*\overset{p}\longrightarrow z^o.$

Combining the uniform convergence of the power with $z^*\overset{p}\to z^o$ gives
$$
|Pow(z^*)-Pow^o(z^o)| \le |Pow(z^*)-Pow^o(z^*)| + |Pow^o(z^*)-Pow^o(z^o)| \overset{p}\longrightarrow0,
$$
where continuity of $Pow^o$ at $z^o$ follows from continuity of the distribution of $Z^o$. Hence
$$
Pow(z^*)\overset{p}\longrightarrow Pow^o(z^o).
$$

For any fixed $m$, conditional on the fitted score function, Theorem~\ref{thm:unec_asymptotic} yields, as $n_2\to\infty$,
$$
\mathrm{Power}_{\mathrm{test}}(\widehat S_{\mathrm{UNEC}}) \longrightarrow Pow(z^*)\{1-(1-\gamma)^m\}.
$$
Taking $n_0,n_1\to\infty$ next therefore gives
$$
\lim_{n_0,n_1\to\infty} \lim_{n_2\to\infty} \mathrm{Power}_{\mathrm{test}}(\widehat S_{\mathrm{UNEC}}) = Pow^o(z^o)\{1-(1-\gamma)^m\}.
$$

By Lemma~\ref{lem:oracle_cutoff}, the finite-$m$ oracle rule is
$S_{z_m^o}^o$, and $z_m^o\longrightarrow z^o$ as $m\to\infty$. Since $Pow^o$ is continuous at $z^o$, $Pow^o(z_m^o)\longrightarrow Pow^o(z^o).$ Therefore,
$$
\begin{aligned}
&\lim_{m\to\infty} \lim_{n_0,n_1\to\infty} \lim_{n_2\to\infty} \left|\mathrm{Power}_{\mathrm{test}}(\widehat S_{\mathrm{UNEC}})-Pow^o(z_m^o)\{1-(1-\gamma)^m\}\right|\\
&\qquad=\lim_{m\to\infty}\{1-(1-\gamma)^m\}\left|Pow^o(z^o)-Pow^o(z_m^o)\right|=0.
\end{aligned}
$$
This proves the claimed asymptotic power optimality.
\end{proof}

\subsection{Proofs for test-specific external thresholds}
\label{app:external_threshold_proofs}

\paragraph{Proof of Theorem~\ref{thm:external_sdr}}
\begin{proof}
Throughout the proof, we work under the conditional law given the training stage and the external target vector $\mathbf c$. Write $\widehat z_j=\widehat Z_{\mathrm{SDR},j}$, with its dependence on $n_2$ understood.

For each fixed $j$, the calibration rule is the common-threshold UNEC procedure with target $c_j$. Theorem~\ref{thm:unec_asymptotic} therefore gives
$$
\widehat z_j\xrightarrow{\mathrm{a.s.}}z_j^*.
$$
As established in the proof of that theorem, $F_j(z_j^*)=\beta_j>0$ and $Q_j(z_j^*)=q$. By continuity of $F_j$ and $Q_j$ at $z_j^*$,
$$
F_j(\widehat z_j)\xrightarrow{\mathrm{a.s.}}\beta_j,
\qquad
Q_j(\widehat z_j)\xrightarrow{\mathrm{a.s.}}q.
$$
Since $m$ is fixed, these convergences hold simultaneously for all $j=1,\ldots,m$. For completeness, set $F_j(+\infty)=0$ and $Q_j(z)=0$ whenever $F_j(z)=0$.

Define the test selection indicators and selection count by
$$
A_j=\mathbbm{1}\{Z(X_{n+j};c_j)\ge\widehat z_j\},
\qquad
R=\sum_{j=1}^m A_j.
$$
Conditional on $\mathcal D_{\mathrm{cal}}$, all cutoffs are fixed, and the test observations remain independent. Consequently,
$$
\mathbb E\left[L_{n+j,j}A_j\mid \mathcal D_{\mathrm{cal}},A_1,\ldots,A_m\right] = A_jQ_j(\widehat z_j).
$$
Taking conditional expectations in the definition of SDR therefore yields
$$
\mathrm{SDR}(\widehat S_{\mathrm{SDR}})=\mathbb E\left[\frac{\sum_{j=1}^m A_jQ_j(\widehat z_j)}{1\vee R}\right].
$$
Using $R/(1\vee R)=\mathbbm{1}\{R>0\}$, we obtain
$$
\begin{aligned}
&\left|\mathrm{SDR}(\widehat S_{\mathrm{SDR}})-q\,\mathbb P(R>0)\right|\\
&\quad\le \mathbb E\left[\frac{\sum_{j=1}^m A_j|Q_j(\widehat z_j)-q|}{1\vee R}\right]\\
&\quad\le \mathbb E\left[\max_{1\le j\le m}|Q_j(\widehat z_j)-q|\right]\longrightarrow 0,
\end{aligned}
$$
where the last step follows from bounded convergence.

Moreover, conditional on $\mathcal D_{\mathrm{cal}}$, the indicators $A_j$ are independent Bernoulli variables with success probabilities $F_j(\widehat z_j)$. Hence,
$$
\mathbb P(R>0)=\mathbb E\left[1-\prod_{j=1}^m \{1-F_j(\widehat z_j)\}\right] \longrightarrow 1-\prod_{j=1}^m(1-\beta_j),
$$
again by bounded convergence. Combining the preceding displays gives
$$
\lim_{n_2\to\infty} \mathrm{SDR}(\widehat S_{\mathrm{SDR}}) = q\left\{1-\prod_{j=1}^m(1-\beta_j)\right\} \le q.
$$
This establishes both the stated limit identity and the asymptotic SDR upper bound.
\end{proof}

\paragraph{Proof of Theorem~\ref{thm:external_power_optimality}}
\begin{proof}
Throughout the proof, the external target vector $\mathbf c$ and the test size $m$ are fixed. Expectations for the data-driven procedure average over the training, calibration, and test samples. Write
$$
s_j^o(x)=\mathbbm{1}\{Z_j^o(x)\ge z_j^o\},
\qquad
\mathbf s^o=(s_1^o,\ldots,s_m^o).
$$
As in the common-threshold analysis, the regularity conditions imply that $z_j^o$ is finite, $F_j^o(z_j^o)>0$, and $Q_j^o(z_j^o)=q$, where $F_j^o(z)=\mathbb P\{Z_j^o(X)\ge z\},$ and $G_j^o(z)=\mathbb P\{Y<c_j,\ Z_j^o(X)\ge z\}.$

For a measurable rule $s:\mathcal X\to\{0,1\}$, define its target-specific power by
$$
\pi_j(s)=\mathbb P\{s(X)=1\mid Y\ge c_j\}.
$$
This is well-defined because $\mathbb P(Y\ge c_j)>0$. Fix any $\mathbf s\in\mathcal A_q$ and a target $j$. If $\mathbb P\{s_j(X)=1\}=0$, then $\pi_j(s_j)=0\le\pi_j(s_j^o)$. A rule selecting with probability one is infeasible because $\mathbb P(Y<c_j)>q$.
Otherwise, Lemma~\ref{lem:oracle_ranking}, applied with $c=c_j$, provides an oracle threshold rule
$$
s_{t_j,j}^o(x)=\mathbbm{1}\{Z_j^o(x)\ge t_j\}
$$
with the same selection probability, no larger conditional false-selection risk, and $\pi_j(s_{t_j,j}^o)\ge\pi_j(s_j)$. Since $s_j$ satisfies the target-specific risk constraint, we have $Q_j^o(t_j)\le q$, thus $t_j\ge z_j^o$ according to its definition. The nesting of oracle threshold sets therefore gives
$$
\pi_j(s_j) \le \pi_j(s_{t_j,j}^o) \le \pi_j(s_j^o), \qquad j=1,\ldots,m.
$$

To translate these inequalities into the test-set power, let denote $H_j=\mathbbm{1}\{Y_{n+j}\ge c_j\},$ and $K=\sum_{j=1}^m H_j.$ Conditional on $(H_1,\ldots,H_m)$, independence of the test observations implies
$$
\mathbb E\left[H_js_j(X_{n+j})\mid H_1,\ldots,H_m\right] = H_j\pi_j(s_j).
$$
Consequently,
$$
\mathrm{Power}\{S(\mathbf s)\} =\mathbb E\!\left[\frac{\sum_{j=1}^m H_j\pi_j(s_j)}{1\vee K}\right] \le\mathbb E\!\left[\frac{\sum_{j=1}^m H_j\pi_j(s_j^o)}{1\vee K}\right] =\mathrm{Power}(\widehat S^o).
$$
Moreover, $\mathbf s^o\in\mathcal A_q$ because $Q_j^o(z_j^o)=q$ for every $j$. Hence
$$
\mathrm{Power}(\widehat S^o) = \sup_{\mathbf s\in\mathcal A_q} \mathrm{Power}\{S(\mathbf s)\}.
$$

Write $\widehat Z_j(x)=\frac{\widehat\mu_{n_0}(x)-c_j}{\widehat\sigma_{n_1}(x)},$ and $\widehat z_j=\widehat Z_{\mathrm{SDR},j}$, and the uniform score-consistency argument in the proof of Theorem~\ref{thm:opt_power} applies separately to each fixed target $c_j$. Since $m$ is fixed, it gives
$$
\Delta_n := \max_{1\le j\le m}\sup_x |\widehat Z_j(x)-Z_j^o(x)| \xrightarrow{p} 0.
$$

Let $\widehat F_{n_2,j}$ and $\widehat G_{n_2,j}$ be the empirical selection and false-selection proportions computed from the fitted scores:
$$
\begin{aligned}
\widehat F_{n_2,j}(z)
&=\frac1{n_2}\sum_{i\in I_{\mathrm{cal}}} \mathbbm{1}\{\widehat Z_j(X_i)\ge z\},\\
\widehat G_{n_2,j}(z)
&=\frac1{n_2}\sum_{i\in I_{\mathrm{cal}}} \mathbbm{1}\{Y_i<c_j\} \mathbbm{1}\{\widehat Z_j(X_i)\ge z\}.
\end{aligned}
$$
Combining the score-sandwich argument in the common-threshold proof with the Glivenko--Cantelli argument in Eq.~\eqref{eq:gc_convergence}, applied to the fixed oracle scores, yields
$$
\max_{1\le j\le m}\left\{\sup_z|\widehat F_{n_2,j}(z)-F_j^o(z)|+\sup_z|\widehat G_{n_2,j}(z)-G_j^o(z)|\right\}\xrightarrow{p}0.
$$
This convergence holds jointly as $n_0,n_1,n_2\to\infty$.

For each fixed $j$, the preceding uniform convergence and Assumption~\ref{assump:opt_power} yield $\widehat z_j\xrightarrow{p}z_j^o$ by the cutoff-consistency argument in the proof of Theorem~\ref{thm:unec_asymptotic}. Possible ties in the fitted scores are covered because uniform convergence to the continuous $F_j^o$ makes the largest empirical score atom $o_p(1)$. Since $m$ is fixed, a finite union bound gives
$$
\max_{1\le j\le m}|\widehat z_j-z_j^o| \xrightarrow{p}0.
$$

Let $A_{n,j}=\mathbbm{1}\{\widehat Z_j(X_{n+j})\ge\widehat z_j\},$ and $A_j^o=\mathbbm{1}\{Z_j^o(X_{n+j})\ge z_j^o\}.$ Set
$$
d_n=\Delta_n+
\max_{1\le j\le m}|\widehat z_j-z_j^o|,
$$
so that $d_n\xrightarrow{p}0$. For any $\varepsilon>0$, on $\{d_n\le\varepsilon\}$, a disagreement between $A_{n,j}$ and $A_j^o$ can occur only if $|Z_j^o(X_{n+j})-z_j^o|\le\varepsilon$. Therefore,
$$
\mathbb P\{\widehat S_{\mathrm{SDR}}\ne\widehat S^o\} \le \mathbb P(d_n>\varepsilon)+\sum_{j=1}^m \mathbb P\{|Z_j^o(X)-z_j^o|\le\varepsilon\}.
$$
First taking the joint sample-size limit and then letting $\varepsilon\downarrow0$, continuity of the oracle score distributions and fixed $m$ imply
$$
\mathbb P\{\widehat S_{\mathrm{SDR}}\ne\widehat S^o\} \longrightarrow 0.
$$
The realized power lies in $[0,1]$ and agrees for the two procedures whenever their selection sets agree. Hence
$$
\left|\mathrm{Power}(\widehat S_{\mathrm{SDR}})-\mathrm{Power}(\widehat S^o)\right| \le \mathbb P\{\widehat S_{\mathrm{SDR}}\ne\widehat S^o\} \longrightarrow 0.
$$
Combining this convergence with oracle optimality proves
$$
\lim_{n_0,n_1,n_2\to\infty} \mathrm{Power}(\widehat S_{\mathrm{SDR}}) = \mathrm{Power}(\widehat S^o) = \sup_{\mathbf s\in\mathcal A_q} \mathrm{Power}\{S(\mathbf s)\}.
$$
\end{proof}

\section{Additional Details for Simulation Studies}
\label{app:simulation_details}

\subsection{Simulation setup and implementation}
\label{app:simulation_setup}


Unless otherwise specified, each simulation uses $n_{\mathrm{train}}=2{,}000$ training observations, $n_{\mathrm{cal}}=1{,}000$ calibration observations, and $n_{\mathrm{test}}=1{,}000$ test observations, with random forests for prediction and scale estimation. We consider target risk levels $q\in\{0.05,0.10,\ldots,0.40\}$ and set $\delta=0$ for UNEC and NP. Within each repetition, all competing methods use the same generated training, calibration, and test samples. Reported risks and powers are averaged over $500$ repetitions for the FDR comparison and the general-loss sensitivity experiments in Appendix~\ref{app:general}, and over $100$ repetitions for the UNEC--SCoRE SDR comparison, the observation-specific threshold experiments, and the two-sided interval experiments.

\paragraph{Mean and scale estimation}
For UNEC, the training sample is equally split into $\mathcal D_{\mathrm{train1}}$ and $\mathcal D_{\mathrm{train2}}$, with $n_{\mathrm{train1}}=n_{\mathrm{train2}}=1{,}000$. A random forest fitted on $\mathcal D_{\mathrm{train1}}$ estimates the conditional mean. On $\mathcal D_{\mathrm{train2}}$, we compute the squared residuals $\{Y_i-\widehat{\mu}(X_i)\}^2$ and fit a second random forest to estimate the conditional variance. The scale estimate $\widehat{\sigma}(x)$ is the square root of the predicted variance.

\paragraph{Loss functions.}
For the one-sided general-loss experiments, we use the sigmoid loss
$$
\ell_\tau(y,c) = \frac{1}{1+\exp\{\tau(y-c)\}}, \qquad \tau>0.
$$
For finite $\tau$, this loss can be viewed as a smooth approximation to the binary false-selection loss, and it converges to $\mathbbm{1}\{y<c\}$ as $\tau\to\infty$ for $y\neq c$. We therefore use $\tau=\infty$ to represent the binary-loss setting.
The interval-target experiments use the bounded distance loss specified in Appendix~\ref{app:simulation_two_sided}.

\paragraph{Comparison methods}
We implement the CS method using the clipped score of \citet{jin2023selection}, which performs best in their numerical experiments. The NP method of \citet{qin2025revamping} uses a Gaussian location-shift working model with $\delta=0$. Both baselines fit their random forest predictors using all $2{,}000$ training observations, as neither CS nor NP implementation requires a separate uncertainty-estimation split.
UNEC-Raw uses the same fitted mean and empirical calibration rule as UNEC but replaces the normalized score with the raw margin $\widehat{\mu}(X)-c$.
These methods are compared under binary false-selection loss, for which SDR reduces to FDR. For each of the five DGPs, we evaluate empirical FDR and selection power across the prespecified target levels. Since CS and NP are designed for binary false-selection control, the subsequent experiments on general MDR and SDR control focus on UNEC.

For general SDR control, we compare UNEC with three variants of SCoRE \citep{bai2026csgeneralrisk}: SCoRE-dtm, SCoRE-homo, and SCoRE-hete. All methods use the sigmoid loss $\ell_\tau(y,c)=\{1+\exp[\tau(y-c)]\}^{-1}$ with $\tau=10$ and the binary reward $r(y,c)=\mathbf{1}\{y>c\}$. For SCoRE, we fit separate random forest regressors to predict loss and reward using all $2{,}000$ training observations, retaining the default hyperparameters used in the authors' implementation. Writing their predictions as $\widehat{L}_\tau(x)$ and $\widehat{R}(x)$, respectively, we use the reward-normalized score
$$
s_q(x)=\frac{\widehat{L}_\tau(x)-q}{\widehat{R}(x)}.
$$
We set SCoRE's parameters to $\alpha=\gamma=q$. The three variants share the same base e-values and differ in the final selection step: SCoRE-dtm applies e-BH directly, whereas SCoRE-homo and SCoRE-hete apply homogeneous and heterogeneous randomized boosting, respectively, before e-BH. We use the efficient e-value computation algorithm corresponding to Algorithm~3 in their paper.
Simulation runtimes include model fitting, prediction, and selection across the full grid of target levels, with fitting and prediction counted once per repetition. Shared computations are charged in full to each SCoRE variant.

\subsection{Data-generating processes}
\label{app:simulation_dgps}

We consider five data-generating processes (DGPs) throughout the experiments, consisting of three location-scale models and two covariate-dependent Gaussian mixture models. DGPs 1--3 follow the location--scale structure used in our power analysis, whereas DGPs 4--5 depart from this structure and are used to assess robustness under more complex conditional distributions.

\paragraph{DGPs 1--3: location--scale models}
Following \citet{jin2023selection}, we generate $\mathbf X=(X_1,\ldots,X_{20})^\top$ with independent coordinates $X_k\sim\mathrm{Unif}(-1,1)$ and
$$
Y=\mu(\mathbf X)+\sigma(\mathbf X)\varepsilon,
\qquad
\varepsilon\sim N(0,1),
$$
where $\varepsilon\sim N(0,1)$ is independent of $\mathbf X$ and
$$
\mu(\mathbf x) = 2\left\{x_1x_2+x_3^2+e^{x_4-1}-1\right\}.
$$
The three DGPs share this conditional mean and differ only in the conditional scale specified in Table~\ref{tab:dgp-location-scale}.

\begin{table}[htbp]
    \centering
    \caption{Conditional scale functions for DGPs 1--3.}
    \label{tab:dgp-location-scale}
    \begin{tabular}{cl}
        \toprule
        DGP & Conditional scale $\sigma(\mathbf x)$ \\
        \midrule
        1 & $0.15$ \\
        \addlinespace
        2 & $0.1\cdot\{5.5-|\mu(\mathbf x)|\}/2$ \\
        \addlinespace
        3 &
        $\begin{aligned}[t]
        0.1\cdot\bigl[0.25\cdot\mu^2(\mathbf x)\mathbbm{1}\{|\mu(\mathbf x)|<2\}+0.5\cdot|\mu(\mathbf x)|\mathbbm{1}\{|\mu(\mathbf x)|\ge1\}\bigr]
        \end{aligned}$ \\
        \bottomrule
    \end{tabular}
\end{table}

DGP 1 is homoscedastic. In DGP 2, the conditional scale decreases with the magnitude of the conditional mean, whereas DGP 3 introduces a piecewise dependence on this magnitude. Together, these settings evaluate selection performance under constant and heterogeneous noise levels.

\paragraph{DGPs 4--5: Gaussian mixture models}
Generate $\mathbf X\sim N(\mathbf 0,I_{20})$ and
$$
Y\mid\mathbf X=\mathbf x \sim \pi(\mathbf x)N\{\mu_1(\mathbf x),\sigma_1^2(\mathbf x)\} + \{1-\pi(\mathbf x)\}N\{\mu_2(\mathbf x),\sigma_2^2(\mathbf x)\}.
$$
The components are specified in Table~\ref{tab:dgp-mixture}, where $\operatorname{expit}(t)=\{1+\exp(-t)\}^{-1}$. DGP 4 introduces mixture components with substantially different variances, while DGP 5 additionally incorporates nonlinear means and covariate-dependent component scales.

\begin{table}[htbp]
    \centering
    \caption{Data-generating mechanisms for DGPs 4 and 5.}
    \label{tab:dgp-mixture}
    \begin{tabular}{lcc}
        \toprule
        Component & DGP 4 & DGP 5 \\
        \midrule
        $\pi(\mathbf x)$
        & $\operatorname{expit}\{4(x_1-0.25)\}$
        & $\operatorname{expit}\{2x_1+x_2x_3-0.5\}$ \\
        $\mu_1(\mathbf x)$
        & $2.7+0.5x_2+0.2x_3^2$
        & $2.7+0.5x_2+0.3x_3^2+0.3x_4x_5$ \\
        $\mu_2(\mathbf x)$
        & $0.7-1.4x_5$
        & $0.7-1.4x_5+0.3\sin(x_6)$ \\
        $\sigma_1(\mathbf x)$
        & $3.0$
        & $2+\operatorname{expit}(x_2+x_3)$ \\
        $\sigma_2(\mathbf x)$
        & $0.30$
        & $0.25+0.20|x_5|$ \\
        \bottomrule
    \end{tabular}
\end{table}

For one-sided experiments with a common target threshold, we set $c=0$ for DGPs 1--3 and $c=1$ for DGPs 4--5.

\subsection{Power versus empirical FDR}
\label{app:fdr_comparison}

\begin{figure}[htbp]
    \centering
    \includegraphics[width=0.6\linewidth]{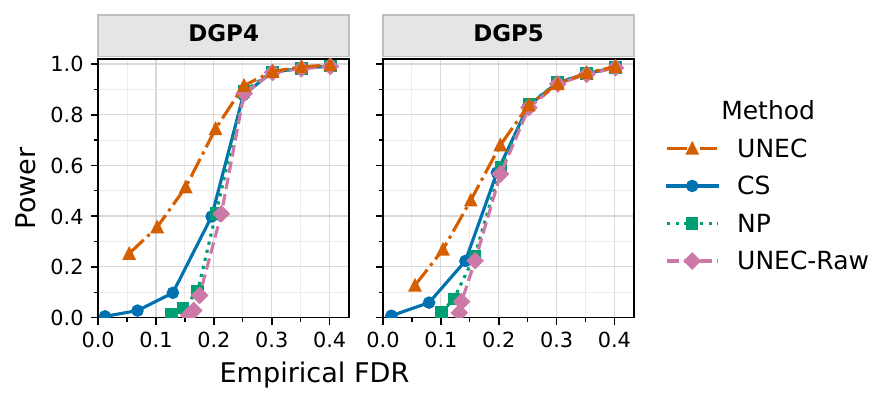}
    \caption{\textbf{Comparison of empirical FDR--power tradeoffs across methods.} Power is presented against the empirical FDR attained by UNEC (triangle dash-dotted red line), CS (circle solid blue line), NP (square dotted green line) and UNEC-Raw (diamond dashed pink line), as the nominal FDR level $q$ varies. The two columns correspond to DGPs 4--5.}
    \label{fig:empirical_fdr_power}
\end{figure}

The nominal-level comparison in Figure~\ref{fig:fdr_power} reflects both differences in ranking and differences in the risk actually attained by each method. To examine these effects, Figure~\ref{fig:empirical_fdr_power} plots power against empirical FDR using the same simulation results.

Figure~\ref{fig:empirical_fdr_power} provides a complementary comparison based on the risk actually attained by each procedure. Under DGPs 4 and 5, UNEC maintains favorable power at comparable empirical FDR levels, indicating that the nominal-$q$ power differences are not explained solely by differential use of the risk budget. Also, the remaining separation between UNEC and UNEC-Raw at comparable empirical FDR is consistent with a benefit from uncertainty-normalized ranking.

\subsection{General-loss results}
\label{app:general}

We next investigate the general risk control of UNEC. Using DGP 5, we consider the sigmoid loss $\ell_\tau(y,c)=\{1+\exp[\tau(y-c)]\}^{-1}$ and fix the target risk level at $q=0.1$. We vary the smoothness parameter $\tau\in\{2,5,10,50,\infty\}$, the calibration sample size $n_{\mathrm{cal}}\in\{100,500,1000,2000\}$, and the adjustment parameter $\delta\in\{0,0.5,1\}$, with $\tau=\infty$ denoting binary loss.

\begin{figure}[ht]
    \centering
    \includegraphics[width=0.85\linewidth]{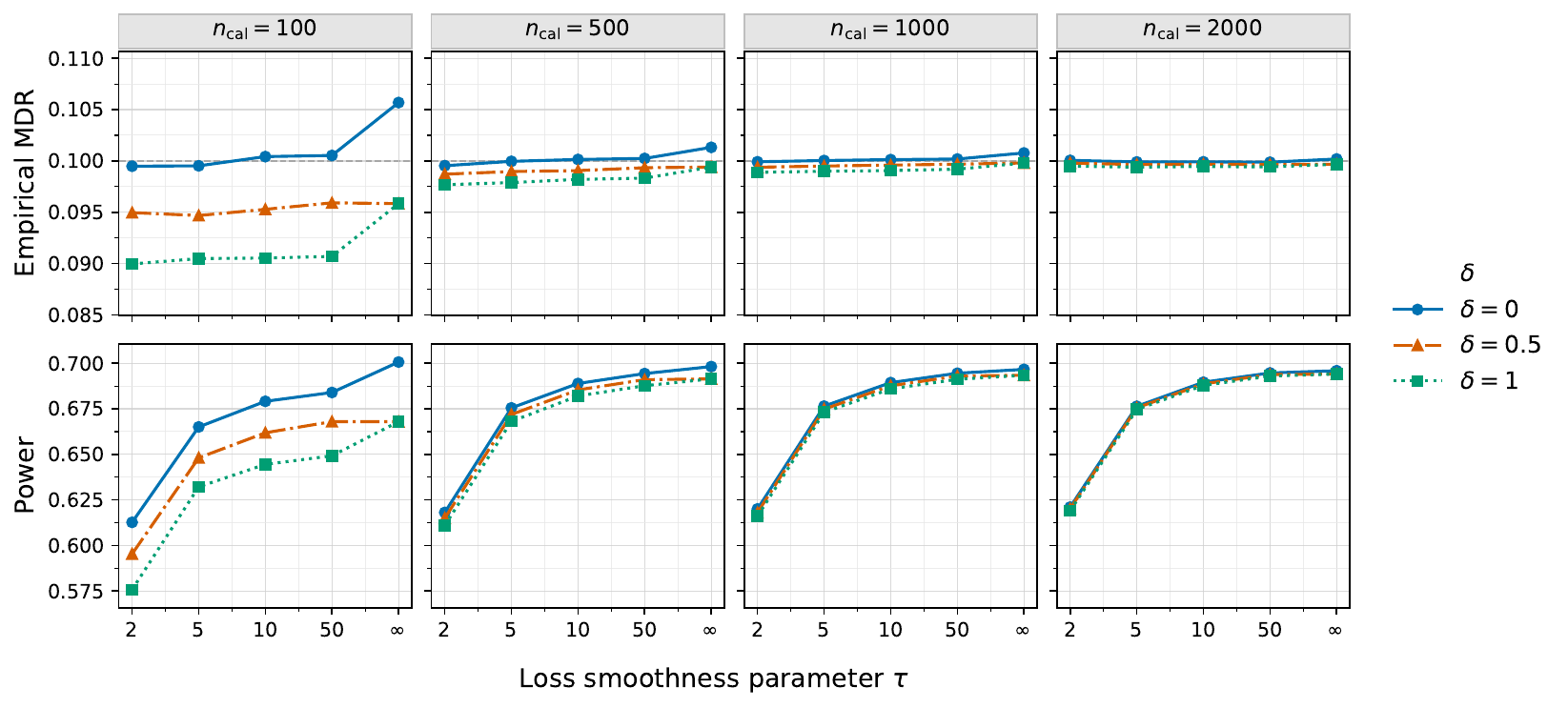}
    \caption{\textbf{MDR control and power under different loss smoothness, calibration sizes, and adjustments.} Empirical MDR and power of UNEC under DGP 5 are presented in the top and bottom rows, respectively, as the loss smoothness parameter $\tau$ varies. The four columns correspond to different calibration sample sizes $n_{\mathrm{cal}}\in\{100,500,1000,2000\}$, and the three curves represent $\delta\in\{0,0.5,1\}$. The target risk level is fixed at $q=0.1$, indicated by the dashed black line in the MDR panels. The case $\tau=\infty$ corresponds to the binary loss.}
    \label{fig:general_mdr}
\end{figure}

\begin{figure}[ht]
    \centering
    \includegraphics[width=0.85\linewidth]{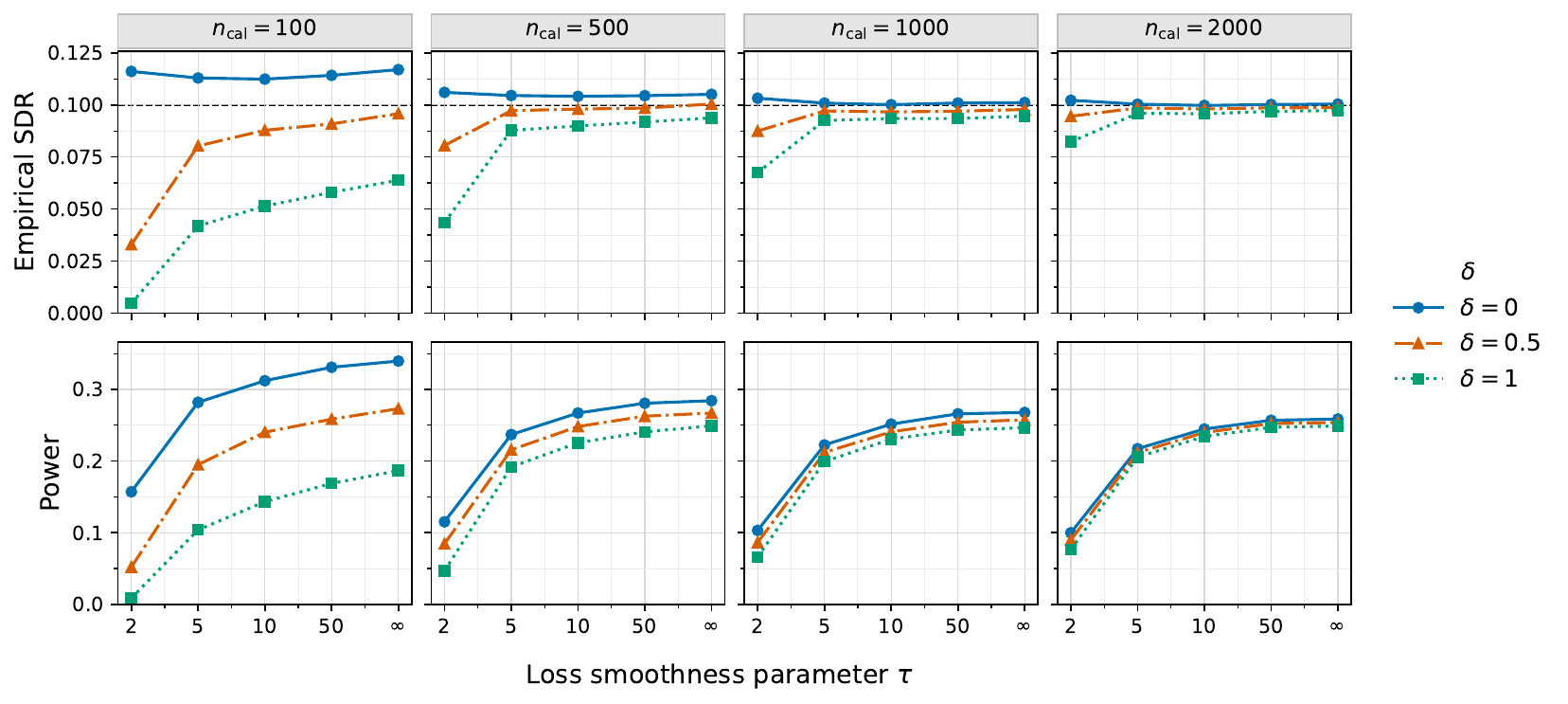}
    \caption{\textbf{SDR control and power under different loss smoothness, calibration sizes, and adjustments.} Empirical SDR and power of UNEC under DGP 5 are presented in the top and bottom rows, respectively, as the loss smoothness parameter $\tau$ varies. The four columns correspond to different calibration sample sizes $n_{\mathrm{cal}}\in\{100,500,1000,2000\}$, and the three curves represent $\delta\in\{0,0.5,1\}$. The target risk level is fixed at $q=0.1$, indicated by the dashed black line in the SDR panels. The case $\tau=\infty$ corresponds to the binary loss.}
    \label{fig:general_sdr}
\end{figure}

Figures~\ref{fig:general_mdr} and~\ref{fig:general_sdr} summarize the performance of UNEC under general bounded losses. Across the considered values of $\tau$, empirical MDR and SDR approaches the target as the calibration size increases, supporting the effectiveness of UNEC under general bounded losses. With $\delta=0$ and $n_{\mathrm{cal}}=100$, the largest empirical MDR is approximately $0.106$, which remains below the finite-sample upper bound $q+\frac{1-q}{n_{\mathrm{cal}}+1}\approx 0.109$ from Corollary~\ref{cor:asymptotic_mdr}. For SDR case, the unadjusted rule can exhibit mild finite-sample exceedance at small $n_{\mathrm{cal}}$, consistent with the asymptotic nature of the SDR guarantee. Generally, positive $\delta$ yields more conservative selections, particularly for small calibration samples, while its effect diminishes as $n_{\mathrm{cal}}$ grows. Power increases with $\tau$ and stabilizes toward the binary-loss case, as the smooth loss increasingly approximates the binary false-selection loss. Overall, these results support the validity of UNEC for general bounded losses while illustrating the finite-sample trade-off induced by the correction parameter $\delta$.

\subsection{Selection with observation-specific thresholds}
\label{app:simulation_varying_thresholds}


We next illustrate the test-specific threshold setting, in which the target threshold varies across test observations. Specifically, for the $j$-th test observation, we consider
$$
H_{0j}:Y_{n+j}<c_j, \qquad H_{1j}:Y_{n+j}\geq c_j, \qquad j=1,\ldots,m.
$$
To introduce different degrees of threshold heterogeneity, we generate $c_j=c_0+2h(U_j-0.5)$, with $U_j\overset{\mathrm{iid}}{\sim}\mathrm{Unif}(0,1)$, independently of the calibration and test data. Equivalently, $c_j\sim\mathrm{Unif}(c_0-h,c_0+h)$, so $h$ is the half-width of the threshold range. We set $c_0=0$ for DGPs 2--3 and $c_0=1$ for DGP 4, and consider $h\in\{0,0.3,0.6,1\}$. Setting $h=0$ recovers the common-threshold setting, while larger $h$ corresponds to greater threshold heterogeneity. We use the sigmoid loss with $\tau=10$ and apply the target-specific calibration procedure in Algorithm~\ref{algo:UNEC_varying}, described in Appendix~\ref{app:varying_thresholds}.

\begin{figure}[ht]
    \centering
    \includegraphics[width=0.7\linewidth]{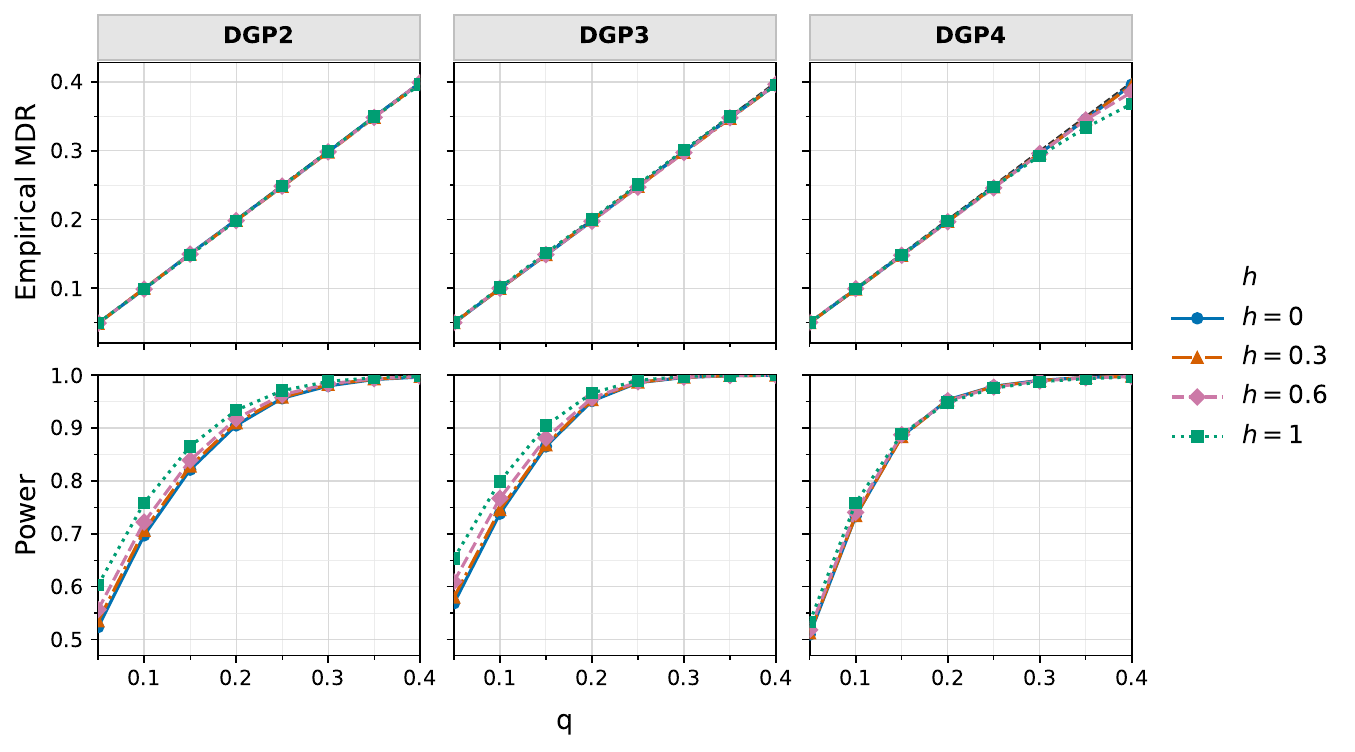}
    \caption{\textbf{MDR control and power with observation-specific thresholds.} Empirical MDR and power of UNEC are presented in the top and bottom rows, respectively, as the target risk level $q$ varies. The three columns correspond to DGPs 2--4, and the four curves represent different degrees of threshold heterogeneity, $h\in\{0,0.3,0.6,1\}$. The dashed black line in the first row is $y=x$.}
    \label{fig:varying_mdr}
\end{figure}

\begin{figure}[ht]
    \centering
    \includegraphics[width=0.7\linewidth]{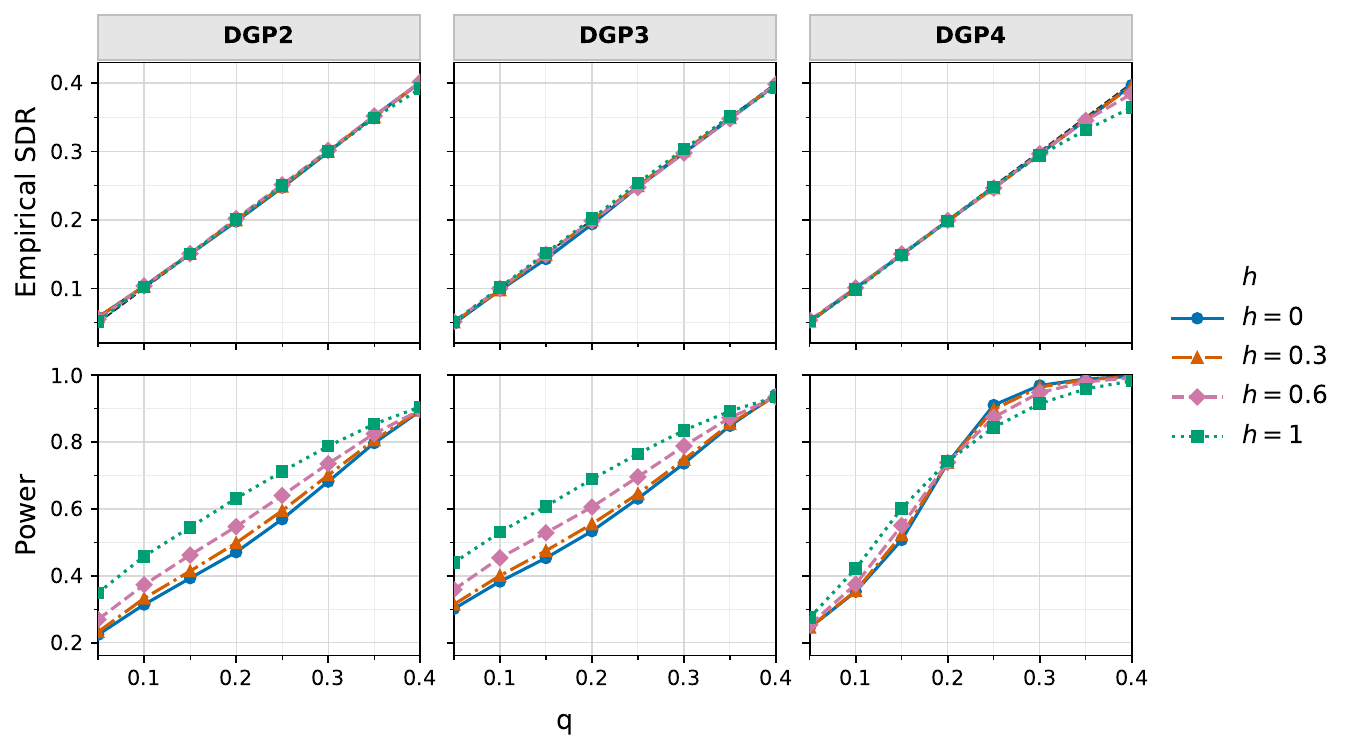}
    \caption{\textbf{SDR control and power with observation-specific thresholds.} Empirical SDR and power of UNEC are presented in the top and bottom rows, respectively, as the target risk level $q$ varies. The three columns correspond to DGPs 2--4, and the four curves represent different degrees of threshold heterogeneity, $h\in\{0,0.3,0.6,1\}$. The dashed black line in the first row is $y=x$.}
    \label{fig:varying_sdr}
\end{figure}

Figures~\ref{fig:varying_mdr} and~\ref{fig:varying_sdr} show that UNEC provides accurate MDR and SDR control across all three DGPs and different levels of threshold heterogeneity. Importantly, increasing $h$ does not noticeably deteriorate risk calibration, indicating that the procedure remains stable as the selection thresholds become more heterogeneous. The threshold heterogeneity has a more visible effect on power. Under DGPs 2 and 3, larger values of $h$ generally lead to higher power, particularly at smaller $q$, whereas the effect is less monotone under DGP 4 and becomes small as $q$ increases. This is expected because changing $h$ alters the distribution of the observation-specific hypotheses and hence their selection difficulty, while the validity of the risk control remains unaffected.

\subsection{Two-sided selection with interval targets}
\label{app:simulation_two_sided}

We consider a two-sided selection problem, where the goal is to identify observations whose responses fall inside target intervals. Specifically, for the $j$-th test observation, we consider
$$
H_{0j}: Y_{n+j}\notin (a_j,b_j), \qquad H_{1j}: Y_{n+j}\in (a_j,b_j), \qquad j=1,\ldots,m.
$$
To define a smooth false-selection loss for this setting, we use the capped distance loss
$$
\ell_\tau(y;a_j,b_j) = \min\left\{1,\,\tau\bigl[(a_j-y)_+ + (y-b_j)_+\bigr]\right\},
$$
where $(x)_+=\max\{x,0\}$. This loss is zero inside the target interval, increases linearly with the distance outside it, and is capped at one. We set $\tau=10$ and use the adjustment $\delta=1$. Throughout this experiment, data are generated from DGP 2. For each test observation, the target interval is constructed as $a_j=M_j-w_0$, and $b_j=M_j+w_0$, where $M_j=m_0+h(U_j-0.5)$, with  $U_j\overset{\mathrm{iid}}{\sim}\mathrm{Unif}(0,1)$ and $m_0=-0.5$. Here, $w_0$ controls the interval half-width, while $h$ controls the heterogeneity of the interval centers. When $h=0$, all intervals are centered at the common value $m_0$, while larger values of $h$ correspond to increasingly heterogeneous interval targets. We consider $w_0\in\{0.7, 0.85, 1\}$ and $h\in\{0, 0.5, 1\}$. The corresponding UNEC calibration algorithm is analogous to that for observation-specific thresholds and is provided in the Appendix~\ref{app:varying_thresholds}.

\begin{figure}[ht]
    \centering
    \includegraphics[width=0.7\linewidth]{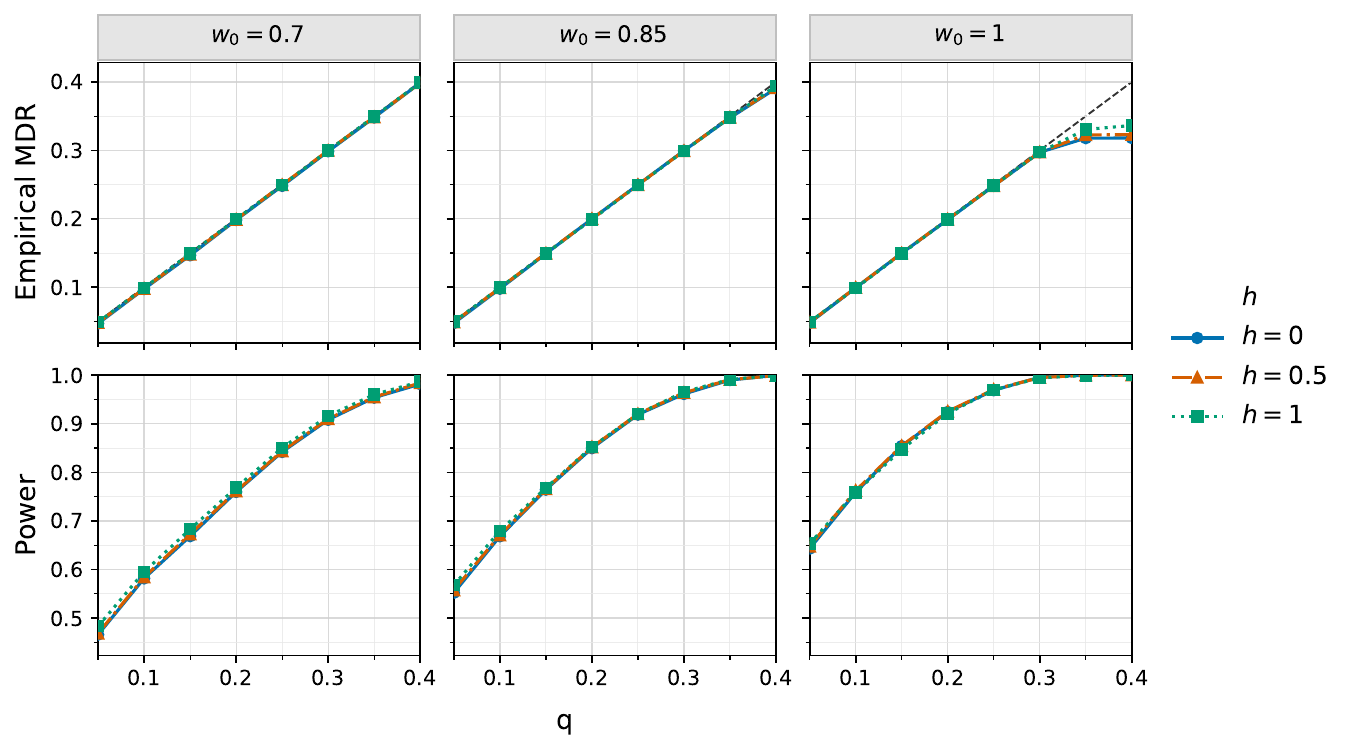}
    \caption{\textbf{MDR control and power for two-sided selection with interval targets.} Empirical MDR and power of UNEC under DGP 2 are presented in the top and bottom rows, respectively, as the target risk level $q$ varies. The three columns correspond to interval half-widths $w_0\in\{0.7, 0.85, 1\}$, and the three curves represent different levels of interval-center heterogeneity, $h\in\{0, 0.5, 1\}$. The capped distance loss is used with $\tau=10$ and adjustment $\delta=1$. The dashed black line in the first row is $y=x$.}
    \label{fig:two_sided_mdr}
\end{figure}

\begin{figure}[ht]
    \centering
    \includegraphics[width=0.7\linewidth]{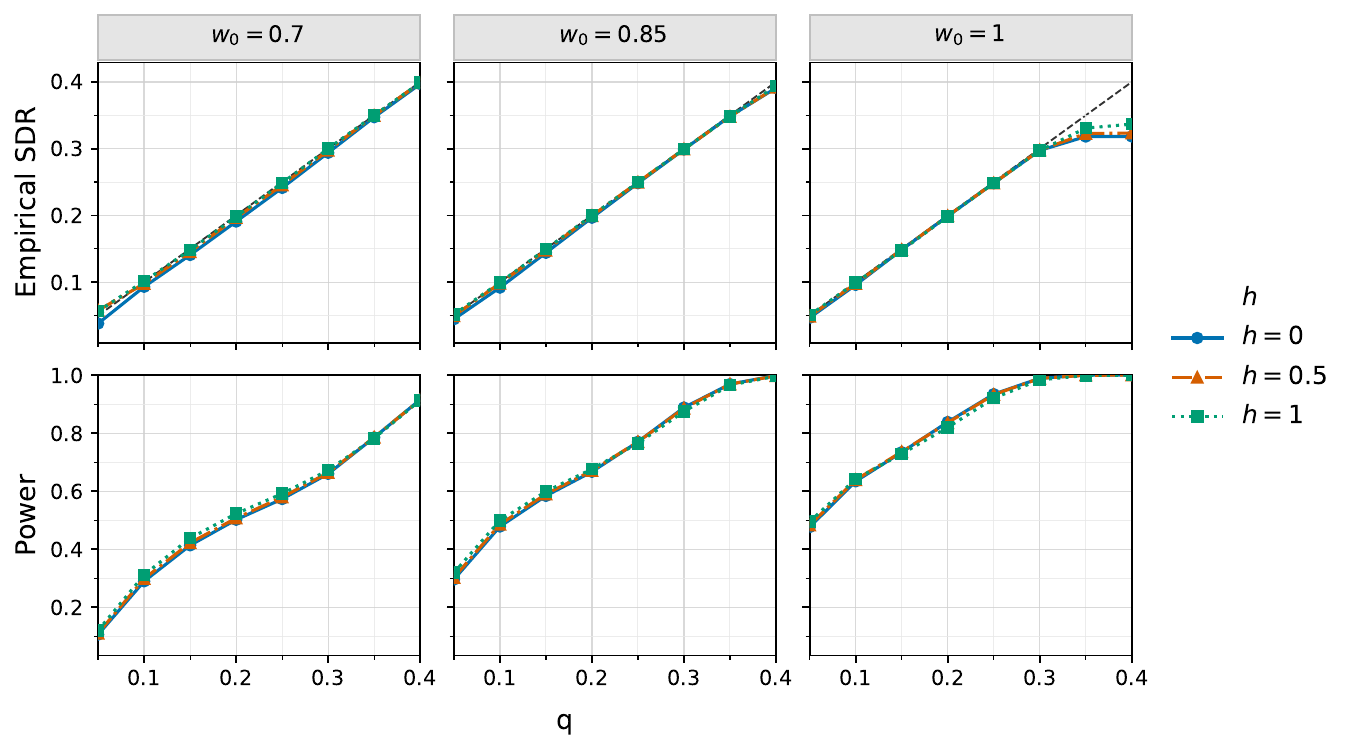}
    \caption{\textbf{SDR control and power for two-sided selection with interval targets.} Empirical SDR and power of UNEC under DGP 2 are presented in the top and bottom rows, respectively, as the target risk level $q$ varies. The three columns correspond to interval half-widths $w_0\in\{0.7, 0.85, 1\}$, and the three curves represent different levels of interval-center heterogeneity, $h\in\{0, 0.5, 1\}$. The capped distance loss is used with $\tau=10$ and adjustment $\delta=1$. The dashed black line in the first row is $y=x$.}
    \label{fig:two_sided_sdr}
\end{figure}

Figures~\ref{fig:two_sided_mdr} and~\ref{fig:two_sided_sdr} show that UNEC provides generally reliable MDR and SDR control in the two-sided setting. Power rises substantially with $w_0$ because wider target intervals make the selection problem easier, while the effect of $h$ is comparatively mild and becomes less pronounced for wider intervals. For $w_0=1$ and large $q$, nearly all test observations are selected, so power approaches one and empirical risk levels off below the nominal target. Overall, these results show that UNEC extends naturally to two-sided interval-target selection and continues to achieve reliable risk control together with strong selection performance.

\section{Additional Details for Drug Discovery Applications}
\label{app:real_data_details}

\subsection{Data processing and model training}
\label{app:implementation}

For each dataset, we allocate 40\% of the observations to training, 10\% to validation, and the remaining 50\% to an evaluation pool. In each of $100$ repetitions, the evaluation pool is randomly split into equally sized calibration and test sets. We evaluate target risk levels $q\in\{0.05,0.10,\ldots,0.40\}$ and report empirical risk, power, and the number of selected compounds, averaged over repetitions. In the AqSolDB FDR comparison, both UNEC and NP use $\delta=0$. UNEC uses $\delta=1$ in the LD50 and lipophilicity experiments, including the additional MDR and SDR results in Appendix~\ref{app:rda_mdr}.

We use two-fold cross-fitting for score estimation. The training sample is divided into two folds, and an AttentiveFP mean model fitted on each fold predicts outcomes in the other fold. These out-of-fold predictions yield absolute residuals for all training observations. We then fit an AttentiveFP scale model to these log-transformed residuals and refit the final mean model on the full training sample. Thus, both final estimators use all available training observations, while each residual is constructed from a mean prediction obtained without using that observation's outcome. The calibration and test samples are excluded from all model-fitting stages. Reported runtimes for the real-data experiments exclude AttentiveFP model training and are averaged over the evaluated target risk levels \(q \in \{0.05, 0.10, \ldots, 0.40\}\).

\subsection{Additional results}
\label{app:rda_mdr}

\paragraph{Toxicity screening with molecule-specific thresholds.}
We report the risk control results for the LD50\_Zhu dataset, using the same screening targets and loss functions as in the main text, while the loss temperature $\tau$ varies in $\{1,2,5,\infty\}$ and adjustment $\delta=1$.

\begin{figure}[ht]
    \centering
    \includegraphics[width=0.85\linewidth]{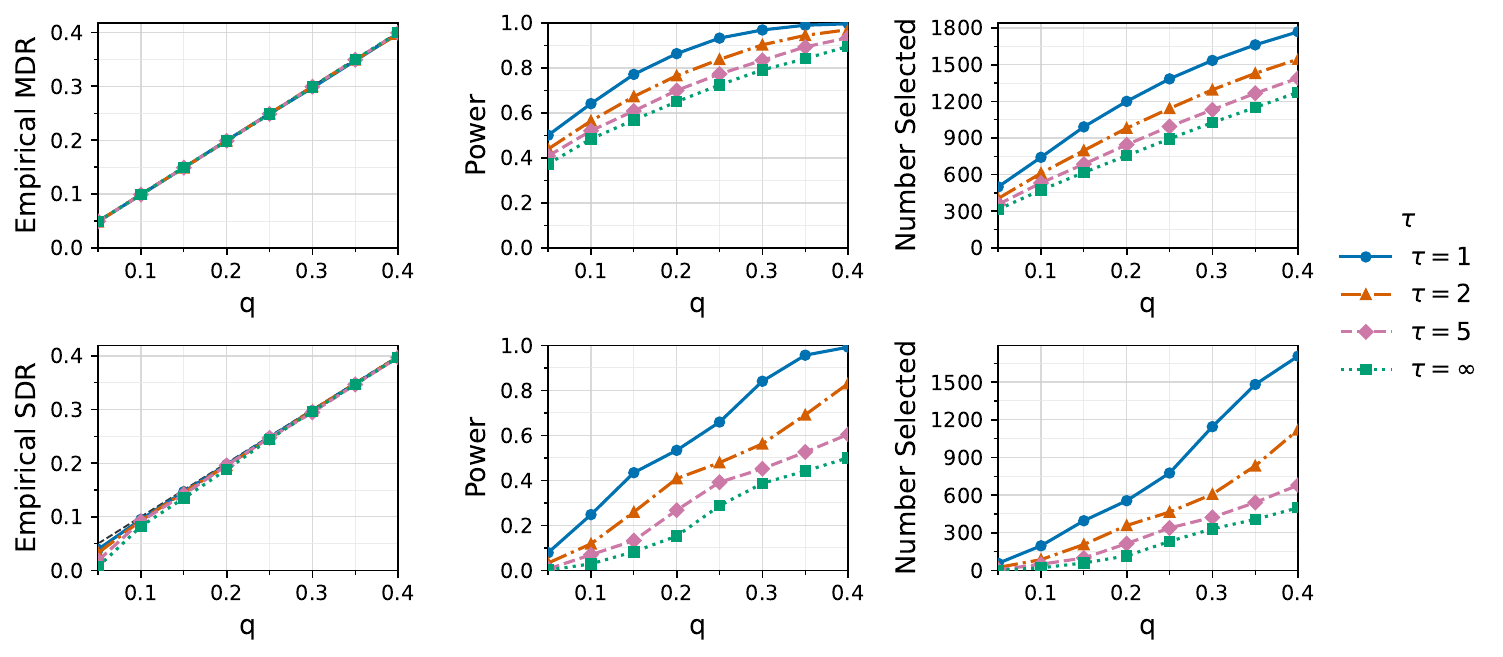}
    \caption{\textbf{Risk control and selection performance for toxicity screening with molecule-specific thresholds.} Empirical MDR/SDR, power, and the average number of selected compounds are presented for UNEC with $\tau=1,2,5,\infty$, where $\tau=\infty$ denotes binary loss. The thresholds $c_j$ correspond to a common $\mathrm{LD}_{50}$ target of $2000$ mg/kg. The dashed black line in the first panel indicates the nominal level $q$.}
    \label{fig:ld50_varying_risk}
\end{figure}

Figure~\ref{fig:ld50_varying_risk} shows that both empirical MDR and SDR are controlled across the four loss settings, although SDR is more conservative at small $q$, whereas MDR tracks the nominal level more closely from below. At a fixed risk level, smaller $\tau$ permits more compounds to be selected and yields higher power by assigning partial penalties to modest shortfalls below the toxicity target. Thus, the loss parameter provides a direct way to regulate the trade-off between the severity of target violations and screening yield.

\paragraph{Two-sided lipophilicity screening.}
We report the corresponding MDR-calibrated results using the same target interval $[1,3]$ and loss functions as in the main text, with adjustment $\delta=1$.

\begin{figure}[ht]
    \centering
    \includegraphics[width=0.85\linewidth]{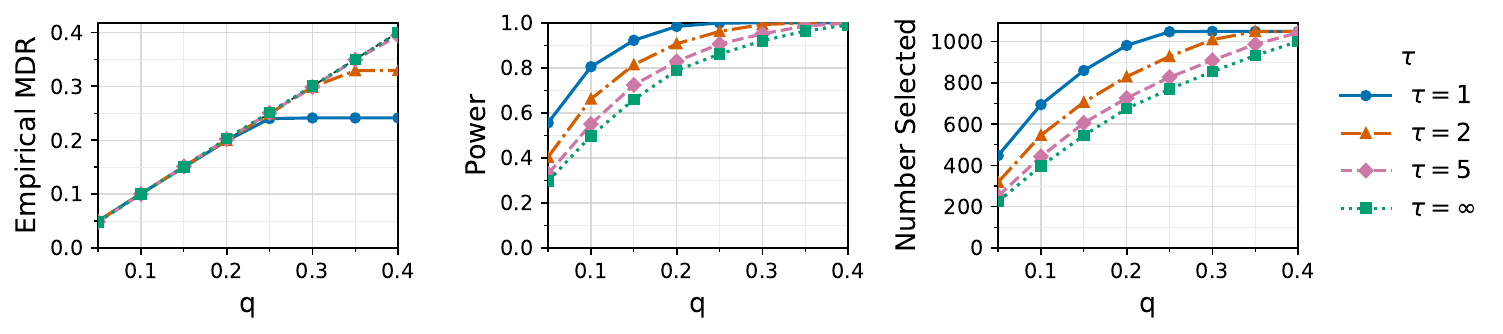}
    \caption{\textbf{MDR control and selection performance for two-sided lipophilicity screening.} Empirical MDR, power, and the average number of selected compounds are shown for UNEC. The target interval is $[1,3]$ on the $\log D_{7.4}$ scale. The four curves correspond to $\tau=1,2,5,\infty$, where $\tau=\infty$ denotes binary loss. The dashed black line in the first panel indicates the nominal level $q$.}
    \label{fig:lipophilicity_two_sided_mdr}
\end{figure}

Figure~\ref{fig:lipophilicity_two_sided_mdr} confirms that empirical MDR generally remains close to or below the nominal level. For smaller $\tau$, its departure below the nominal curve at larger $q$ coincides with selection approaching the full test set. When all test compounds are selected, empirical MDR and SDR both equal the average loss over that set, explaining the similar plateaus under the two calibration criteria.
\end{document}